\pdfoutput=1
\documentclass[12pt]{article}

\font\grande=cmr9.5 scaled \magstep4
\font\medio=cmr9.5 scaled \magstep2
\outer\def\beginsection#1\par{\medbreak\bigskip
      \message{#1}\leftline{\bf#1}\nobreak\medskip
\vskip-\parskip
      \noindent}
\usepackage{graphicx} 
\begin{document}
\bibliographystyle {unsrt}

\titlepage

\begin{flushright}
CERN-TH-2016-193
\end{flushright}

\vspace{15mm}
\begin{center}
{\grande Spectator Higgs, large-scale gauge fields}\\
\vspace{5mm}
{\grande and the non-minimal coupling to gravity}\\
\vspace{15mm}
 Massimo Giovannini 
 \footnote{Electronic address: massimo.giovannini@cern.ch} \\
\vspace{0.5cm}
{{\sl Department of Physics, Theory Division, CERN, 1211 Geneva 23, Switzerland }}\\
\vspace{1cm}
{{\sl INFN, Section of Milan-Bicocca, 20126 Milan, Italy}}
\vspace*{1cm}

\end{center}

\centerline{\medio  Abstract}
\vskip 0.5cm
Even if the Higgs field does not affect the evolution of the background geometry, 
its massive inhomogeneities induce large-scale gauge fields whose energy density 
depends on the slow-roll parameters, on the effective scalar mass and, last but not least, 
on the dimensionless coupling to the space-time curvature. 
Since the non-Abelian gauge modes are screened, 
the non-minimal coupling to gravity predominantly affects the evolution of the 
hypercharge and electromagnetic fields. While in the case of minimal coupling the obtained constraints are immaterial, 
as soon as the coupling increases beyond one fourth the produced fields become 
overcritical. We chart the whole parameter space of this qualitatively new set of bounds.
Whenever the limits on the curvature coupling are enforced, the magnetic field
may still be partially relevant for large-scale magnetogenesis and 
exceed  $10^{-20}$ G for the benchmark scale of the protogalactic collapse.

\noindent

\vspace{5mm}

\vfill
\newpage

\renewcommand{\theequation}{1.\arabic{equation}}
\setcounter{equation}{0}
\section{Introduction} 
\label{sec1}
According to the current evidence firstly established by the WMAP experiment \cite{WMAP1}, 
the observed temperature and polarization anisotropies are consistent with the 
theoretical expectations iff the initial conditions of the Einstein-Boltzmann hierarchy are predominantly 
adiabatic and Gaussian (see also \cite{WMAP2,ACBARQUAD}). 
Every deviation from this concordance paradigm leads to entropic initial data (see e.g.\cite{hh}).  
When the Cosmic Microwave Background (CMB) experiments are combined 
with the two remaining sets of cosmological observations \cite{LSS,SNN} the parameters describing the 
large-scale curvature modes (e.g. spectral index, normalization amplitude, tensor to scalar ratio) are 
slightly (but not crucially) affected.  The adiabatic lore is then compatible with a minute tensor to scalar ratio $r_{T}$ 
provided some kind of plateau-like potential  dominates, even before the onset of the inflationary stage, 
against the kinetic energy of the inflaton and against the spatial curvature \cite{WMAP2}.

Even assuming that the energy density of the Higgs field with typical mass  ${\mathcal O}(125) \, \mathrm{GeV}$  
\cite{HG} is always subdominant during the conventional inflationary evolution \cite{k1}, the inhomogeneities of 
any spectator field are amplified during inflation and contribute, ultimately, to the total curvature budget. 
When the inflationary curvature scale\footnote{The following notations for the Planck mass will be used throughout: 
$\overline{M}_{P} = 1/\sqrt{8\pi G} = M_{P}/\sqrt{8 \pi}$ where $M_{P} = 1.22 \times 10^{19} \, \mathrm{GeV}$.} $H_{e}$  is much 
smaller than $10^{-6} \,M_{P}$ \cite{low} nearly all the curvature perturbations observed in the microwave sky 
might plausibly come from the quasi-flat spectrum of the Higgs field. While general arguments suggest that the inflationary 
rate could be lowered between $10^{-12} \, M_{P}$ and $10^{-7}\,M_{P}$ (see for instance \cite{low}), this possibility 
does not apply to the specific case of the Higgs. The lower bound on the expansion rate
(i.e. $H_{e} \geq 10^{-12} \, M_{P}$) depends actually on the form of the post-inflationary potential.
If the quartic term dominates the potential after inflation  the corresponding energy density evolves in average as 
$a^{-4}$ \cite{tur} where $a$ denotes the scale factor of a 
Friedmann-Robertson-Walker metric\footnote{In the present paper the background metric will always be considered conformally flat
and denoted by $\overline{g}_{\mu\nu} = a^2(\tau) \, \eta_{\mu\nu}$ where $\eta_{\mu\nu}$ is the Minkowski metric and 
$a(\tau)$ is the scale factor typically expressed as a function of the conformal time coordinate.}. 
Consequently, if the Universe reheats suddenly after the end of inflation the Higgs energy density 
will be negligible. Similarly the Higgs inhomogeneities will be much smaller than the observed value of large-scale curvature 
perturbations  \cite{ch}. 

The conclusion of the previous paragraph is not specific to the Higgs but it depends on the 
shape of the potential and on the post-inflationary dynamics of the radiation plasma.
If the evolution after inflation is instead dominated by a fluid with equation of state stiffer than radiation the ratio 
of the Higgs field to the background can potentially increase and the induced 
curvature inhomogeneities might get larger. Unconventional post-inflationary evolutions of this type 
have been discussed long ago  \cite{low} (see also \cite{low2} for the gravitational waves produced in this context). 
The inflationary fluctuations of the Higgs field might also modulate the reheating process \cite{mod}.
However, since it is reasonable to expect that the Higgs will decay before 
becoming dominant, the induced curvature inhomogeneities could be excessively non-Gaussian, as implied 
by general arguments related to the dynamics of the spectator fields \cite{low}. This last class of scenarios 
turns out to be strongly constrained by the observational limits on non-Gaussianities \cite{WMAP1,WMAP2}.

An explicit (non-minimal) coupling to the curvature 
may lead to a significant production of Higgs particles \cite{rec} which have been taken to be massless both during 
inflation and in the subsequent radiation-dominated phase (see also \cite{SO} for earlier discussions on this issue).  The contribution of the Higgs 
mass can be safely neglected all along the inflationary phase but it becomes important  
during the radiation epoch. Another potential limitation involves the gauge fields. 
It is well established that the evolution of the inhomogeneities of the 
spectator fields amplifies the gauge fields \cite{mg0,mg2}. It is then interesting to include the coupling of the Higgs to gravity
and to compute the induced large-scale magnetic fields.
Qualitatively new bounds on the coupling of the Higgs to the space-time curvature will emerge from these considerations. In specific 
corners of the parameter space the produced gauge fields will turn out to be phenomenologically relevant (for an introduction 
to the problem of magnetogenesis see, for instance, \cite{mg3}).

On a general ground, large-scale magnetic fields produced during inflation can affect various phenomena and, in particular, 
galactic magnetism. While the correlation scale of the field must be sufficiently large (probably exceeding the Mpc 
at the onset of the rotation of the protogalaxy), its amplitude must not jeopardize the closure bound for all the typical scales of the problem \cite{mg0,mg3}. 
The non-minimal coupling to the space-time curvature (parametrized by the 
dimensionless constant $\xi$) increases the produced magnetic field for  comoving scales of the order of the Mpc
 but it might also saturate the closure bound at smaller distance scales. It is then a quantitative issue to chart the corners of the 
parameter space where sufficiently strong magnetic fields may seed either the galactic dynamo or even the compressional 
amplification alone \cite{mg3} (see also \cite{parker} for a classic treatise on these themes). This dual analysis, to the best of our knowledge, has not been 
attempted before even if there are available results in the minimally coupled case \cite{mg2}. We shall show that the more general approach 
of this paper is fully compatible with that previous results that can be accurately recovered in the limit $\xi \to 0$.
Similarly the obtained limits will be immaterial in the conformally coupled case corresponding, within the present conventions, to the 
value $\xi \to -1/6$.

 In short the logic and the main purpose of the present investigation are the following. The produced gauge fields depend, among other things, 
on the specific coupling of the Higgs field to the scalar curvature. While in the case of minimal coupling (i.e. $\xi =0$) the constraints 
are negligible (see also \cite{mg2}), as soon as the coupling increases the produced fields may saturate and even exceed the critical density bound.
The main purpose will then be to chart the parameter space of the model with the aim of constraining the Higgs coupling to the 
space-time curvature. It will also be interesting to scrutinize more closely those regions regions where 
the fields are not overcritical but can still be relevant for the problem of magnetogenesis \cite{mg3}.
The layout of the paper can be summarized as follows. In section \ref{sec2}  we address the classical and quantum evolution 
of the Higgs inhomogeneities by including the coupling to the space-time curvature. In section \ref{sec3}  we shall compute the 
production of the massive modes and of the corresponding currents. The magnetic fields and the bounds on the curvature coupling 
will be discussed in section \ref{sec4}. Section \ref{sec5} contains the concluding remarks.  
\renewcommand{\theequation}{2.\arabic{equation}}
\setcounter{equation}{0}
\section{Higgs inhomogeneities} 
\label{sec2}
The Higgs sector of the standard model action in a four-dimensional curved background can be written as: 
\begin{equation}
S = \int d^{4} x \sqrt{-g} \biggl[ - \frac{R}{ 2 \ell_{P}^2}  + | D_{\mu}\hat{H}|^2 - V (|\hat{H}|) - \xi R |\hat{H}|^2 \biggr],
\label{first}
\end{equation}
where $\ell_{P}= 1/\overline{M}_{P}$, $D_{\mu}$ is the $SU_{L}(2)\otimes U_{Y}(1)$ covariant derivative and $V(|\hat{H}|)$ is the 
Higgs potential; the non-minimal coupling has been denoted by $\xi$ and the complex Higgs doublet $\hat{H}$ is given
 in terms of four real scalar fields. In curved backgrounds the production of scalar particles in general (and of Higgs particles in particular) can be studied 
 with various methods and within different approximations. The simplest approach is to disregard completely the mass and the interactions with the gauge sector. 
 The particles produced when $\xi\neq 0$ can be computed analytically and this strategy has been followed, for 
 instance, in Ref. \cite{rec}. Whenever $\xi$ is sufficiently close to the conformal case it is even possible to tailor a perturbative expansion whose small 
 parameter is exactly the deviation of $\xi$ from its conformal value \cite{staro,birrel,ford1}. In the case of minimal coupling $\xi$ is exactly equal to zero and 
 the conformal expansion is reasonably well defined (see for instance the second paper in \cite{ford1} and discussion therein). 
 
The non-Abelian gauge fields are screened as the Universe thermalizes and this 
phenomenon has been investigated for the first time in connection with chromo-electric 
and chromo-magnetic fields at finite temperature \cite{chrom}.
Conversely the hypercharge fields are not screened and remain unscreened 
after the electroweak phase transition \cite{mg0,mg2}. Since the contribution of the Abelian fields 
cannot be overlooked it seems rather plausible to scrutinize their amplification 
and their potential effects. We shall then analyze the gauge fields induced by the massive 
Higgs fluctuations in the framework of the Abelian-Higgs model. While the mass does not affect the 
mode functions during inflation, it becomes essential in the radiation epoch. 
Conversely, in the case of a spatially flat geometry (such as the 
one suggested by current observations) the dimensionless parameter $\xi$ 
drops out from the evolution equations of the mode functions during 
radiation. This happens because the Ricci scalar vanishes exactly on the background.
\subsection{Abelian-Higgs model during inflation} 
The total action of the Abelian-Higgs model non-minimally coupled to gravity can be expressed as:
\begin{equation}
S = \int d^4 x \sqrt{-g} \biggl[  - \frac{R}{ 2 \ell_{P}^2}  + 
g^{\mu\nu}\, ({\mathcal D}_{\mu} \phi)^{\ast} {\mathcal D}_{\nu} \phi - V(|\phi|) - \xi R \phi^{\ast} \phi
-\frac{1}{4}  Y_{\alpha\beta} Y^{\alpha\beta} \biggr],
\label{AH1}
\end{equation}
where ${\mathcal D}_{\mu} = \nabla_{\mu} + i q  Y_{\mu}$ is the $U(1)$ covariant derivative, $\nabla_{\mu}$ is the generally covariant 
derivative; $ -Y_{\alpha\beta} Y^{\alpha\beta}/4 $ denotes the standard kinetic term of the gauge field. To avoid 
potential confusions we want to stress that $\phi$ is not the inflaton field but rather the Higgs field of the Abelian-Higgs model. In this 
paper we shall study the evolution of the fluctuations of the Higgs field in a fixed inflationary background 
described in terms of its slow-roll dynamics. The evolution equations 
obtained from Eq. (\ref{AH1}) are (see also \cite{ford2}):
\begin{eqnarray}
&& g^{\alpha\beta} {\mathcal D}_{\alpha} {\mathcal D}_{\beta} \phi + \frac{\partial V}{\partial \phi^{*}}  + \xi R \phi=0,
\label{AH1a}\\
&& \nabla_{\mu} Y^{\mu\nu} = j^{\nu} - 2 q^2 Y^{\nu} \phi^{*} \phi,
\label{AH1b}
\end{eqnarray}
where the current appearing in Eq. (\ref{AH1b}) is given by:
\begin{equation}
j^{\nu} = g^{\mu\nu} j_{\mu}, \qquad j_{\mu} = i q  [ \phi^{*} \partial_{\mu} \phi - \phi \partial_{\mu} \phi^{*}].
\label{AH1d}
\end{equation}
Equations (\ref{AH1a}) and (\ref{AH1b})  are supplemented by the Einstein equations which are not affected by the Higgs both during and after inflation.
During inflation the total energy-momentum tensor is dominated by the contribution of the inflaton whose energy density 
greatly exceeds the one of the Higgs boson. After inflation the energy density of the radiation background 
will be the leading component of the energy-momentum tensor of the plasma 
even if different possibilities are not excluded by the present analysis (see, for instance, Ref. \cite{low} and discussion therein). 
As it will be clear in section \ref{sec3} we shall work within the sudden reheating approximation where the scale factor and the extrinsic curvature 
are continuous across the inflationary boundary.
Since the Higgs field does not dominate at any stage of the evolution of the background it is natural to 
posit that $\xi \overline{R} \phi^{*} \phi \ll R \overline{M}_{P}^2$ where 
$\overline{R}$ denotes the background Ricci scalar. The previous requirement  guarantees that the effects of the amplified inhomogeneities 
on the total curvature $R$ (i.e. background plus fluctuations) can also be ignored. In this approach the energy density of the Higgs (or of the gauge fields)
must never exceed the critical energy density. 

We consider the situation where the background gauge fields are vanishing, i.e. $\overline{Y}_{\alpha} = \overline{Y}_{\alpha\beta} =0$
but their fluctuations can be dynamically generated thanks to the currents of the massive Higgs inhomogeneities \cite{mg2}. To lowest order the evolution equation of 
$\phi$ is only affected by $\xi$ and can be written, in general terms,  as 
\begin{equation}
\overline{g}^{\alpha\beta} \, \overline{\nabla}_{\alpha} \, \overline{\nabla}_{\beta} \phi + m^2 \phi+ \xi \overline{R} \phi=0,
\label{AH2}
\end{equation}
where $\overline{\nabla}_{\alpha}\, \overline{\nabla}_{\beta} = \partial_{\alpha}\partial_{\beta} - \overline{\Gamma}_{\alpha\beta}^{\mu} \partial_{\mu}$
and $\overline{\Gamma}_{\alpha\beta}^{\mu}$ denotes the background Christoffel symbols. 
For the sake of accuracy we mention that, within the present conventions, the Riemann tensor 
is given by $R^{\beta}_{\,\,\mu\alpha\nu} = \partial_{\alpha} \Gamma^{\beta}_{\mu\nu} +\,.\,.\,.$; the Ricci tensor is defined from the 
contraction of the first and third indices of the Riemann tensor, i.e. $R_{\mu\nu} = R^{\alpha}_{\,\,\,\mu\alpha\nu}$.
The Ricci scalar of the background (in the case of a conformally flat metric $\overline{g}_{\mu\nu} = a^2 \eta_{\mu\nu}$ 
with signature $(+,\, -,\,-,\,-)$) is then given by $\overline{R} = - 6 a^{\prime\prime}/a^3$. With these
specifications, Eq. (\ref{AH2}) becomes:
\begin{equation}
\Phi''  - \nabla^2 \Phi - \frac{a''}{a} \Phi + m^2 a^2  \Phi - 6 \xi \frac{a^{\prime\prime}}{a} \Phi=0,\qquad \Phi = a \phi,
\label{AH3}
\end{equation}
and analogously for the complex conjugate field. In Eqs. (\ref{AH2}) and (\ref{AH3}) $m$ denotes the effective mass term and does not only necessarily coincide with the 
 the mass term of the potential, (i.e. $m_{0}$) but it can also include the contribution of the quartic term. Denoting with $v$ the vacuum expectation value of $|\phi|$ during inflation, whenever $v \gg {\mathcal O}(250)$ GeV,  $m_{0}$ can be negligible if compared with the contribution of quartic self-interaction which is of the order of $\lambda v^2$. 
 
We would like to stress that the background metric will be taken to be conformally flat. There are various reasons for this choice: first any spatial 
curvature will be exponentially suppressed during inflation; second the concordance scenario stipulates that today the contribution 
of the spatial curvature to the energy budget of the Universe is negligible meaning that we are, to a good approximation, rather close 
to the critical density. During inflation, however, the metric fluctuations are also amplified and the interplay between the 
metric fluctuations and the Higgs inhomogeneities is negligible. After inflation the Higgs inhomogeneities might affect the large-scale 
curvature inhomogeneities as it happens in the case of spectator fields of different nature. This however, does not happen.
As explained in the introduction, the energy density of the zero mode of the Higgs field evolves like $a^{-4}$. Thus during 
radiation the amplified Higgs inhomogeneities will remain small and will leave unaffected the temperature and polarization 
anisotropies of the Cosmic Microwave Background \cite{ch}.
 
What matters is the ratio between $m$ and the expansion rate both during inflation and in the subsequent radiation-dominated phase. The effective mass will then enter the (four-dimensional)  parameter space through the quantity $\zeta = m/(a_{e} H_{e})$ where $H_{e}$ is the curvature scale at the end of inflation and $a_{e}$
 is the scale factor at the corresponding epoch. Recall, in this respect, that the inflationary expansion rate and the inflaton potential 
 in Planck units can be expressed as:
 \begin{eqnarray}
\biggl(\frac{H_{e}}{M_{P}}\biggr) &=& 6.85\times 10^{-6} \,\biggl(\frac{r_{T}}{0.1}\biggr)^{1/2} \biggl(\frac{{\mathcal A}_{{\mathcal R}}}{2.4\times 10^{-9}}\biggr)^{1/2},
\label{sc1}\\
\biggl(\frac{W}{M_{P}^4}\biggr) &=& 5.6 \times 10^{-12} \, \biggl(\frac{r_{T}}{0.1}\biggr) \biggl(\frac{{\mathcal A}_{{\mathcal R}}}{2.4\times 10^{-9}}\biggr),
\label{sc2}
\end{eqnarray}
where ${\mathcal A}_{{\mathcal R}}$ the amplitude of the scalar power spectrum
 at the conventional pivot wavenumber $k_{p}=0.002\,\, \mathrm{Mpc}^{-1}$; $W$ denotes the inflaton potential. 
If the consistency relations are enforced (as assumed in Eqs. (\ref{sc1}) and (\ref{sc2})) 
 the tensor to scalar ratio $r_{T}$, the tensor spectral index $n_{T}$ and the slow roll parameter $\epsilon$ obey the following 
chain of equalities $r_{T} = {\mathcal A}_{T}/{\mathcal A}_{{\mathcal R}} = 16\epsilon = - 8 n_{T}$ 
where $\epsilon = - \dot{H}/H^2$ is the slow-roll parameter and ${\mathcal A}_{T}$ is the amplitude of the tensor power spectrum at the 
same pivot scale $k_{p}$ used to assign the scalar power spectrum.
All in all since the value of the quartic self-interaction $\lambda$ is ${\mathcal O}(10^{-2})$ 
we can say, with a fair degree of confidence, that the condition $m/H_{e} \ll 1$ will always be verified in practice,

Let us now conclude this discussion by mentioning that, in Fourier space, Eq. (\ref{AH3}) becomes: 
\begin{equation}
\Phi_{\vec{k}}^{\prime\prime} + \biggl[ k^2 + m^2 a^2 - (1 + 6 \xi) \frac{a^{\prime\prime}}{a} \biggr] \Phi_{\vec{k}} =0.
\label{AH4}
\end{equation}
In terms of $\Phi$ the total (rescaled) current of Eq. (\ref{AH1d}) can be expressed as:
\begin{equation}
J^{\mu} = \sqrt{-g} \, g^{\mu\nu} \,j_{\nu} = i \,q\,  \eta^{\mu\nu}\, [ \Phi^{*} \partial_{\nu} \Phi - \Phi \partial_{\nu} \Phi^{*}],
\label{AH4a}
\end{equation}
where $\eta_{\mu\nu}$ denotes the Minkowski metric. In Eqs. (\ref{AH3}) and (\ref{AH4})  the cases of conformal and minimal couplings 
correspond, respectively,  to $\xi \to -1/6$ and $\xi\to 0$. Note that the conformal coupling corresponds 
to $\xi \to + 1/6$ in Refs. \cite{birrel,ford1} where the Ricci scalar of the background has an opposite sign due to the different conventions 
defining the Riemann tensor. See also, in this respect, the discussion after Eq. (\ref{AH3}).

 \subsection{Action for the Higgs inhomogeneities on a fixed background}
From now on the attention will be focussed on the case of a conformally flat background geometry (i.e. $\overline{g}_{\mu\nu}(\tau) = a^2(\tau) \eta_{\mu\nu}$).
The action for the rescaled field $\Phi$ can be obtained from Eq. (\ref{AH1}) and the explicit result is:
\begin{eqnarray}
S &=& \int d^{3} x \, \int d\tau \biggl[\Phi^{*\,\prime} \Phi^{\prime} + {\mathcal H}^2 \Phi^{*} \Phi - {\mathcal H} (\Phi^{*} \Phi^{\prime} + \Phi \Phi^{*\,\prime})
\nonumber\\
&-& m^2 a^2 \Phi^{*} \Phi + 6 \xi ( {\mathcal H}^2 + {\mathcal H}^{\prime}) \Phi^{*} \Phi - \partial_{i} \Phi^{*} \partial^{i} \Phi\biggr],
\label{AH5}
\end{eqnarray}
where ${\mathcal H} = a^{\prime}/a$ and the prime denotes a derivation with respect to $\tau$. 
For the action explicitly depends on time, non-covariant total derivatives can be added or subtracted 
without affecting the evolution equations of $\Phi$. The different results obtained in this way will lead to slightly 
different Hamiltonians all related by (time-dependent) canonical transformations. Indeed recalling the obvious relation 
$( {\mathcal H} \Phi^{*} \, \Phi)^{\prime} = {\mathcal H}^{\prime} \Phi^{*} \, \Phi +  {\mathcal H} (\Phi^{*} \Phi^{\prime} + \Phi \Phi^{*\,\prime})$,
each of the two terms at the right hand side of this equation can be alternatively 
eliminated from Eq. (\ref{AH5}). Consequently two complementary forms of the Hamiltonian are given by:
\begin{eqnarray}
H_{1}(\tau) &=& \int d^{3} x \biggl[ \Pi_{1}^{*} \Pi_{1} + \partial_{i} \Phi^{*} \partial^{i} \Phi + m^2 a^2 \Phi^{*} \Phi 
\nonumber\\
&+& {\mathcal H}( 1 + 6 \xi) ( \Phi^{*} \Pi_{1}^{*} + \Phi \Pi_{1}) + 6 \xi (6 \xi + 1) {\mathcal H}^2 \Phi^{*}\Phi \biggr], 
\label{AH5b}\\
H_{2}(\tau) &=& \int d^{3} x \biggl[ \Pi_{2}^{*} \Pi_{2} + \partial_{i} \Phi^{*} \partial^{i} \Phi + m^2 a^2 \Phi^{*} \Phi - (6 \xi + 1) ( {\mathcal H}^2 + {\mathcal H}^{\prime}) \Phi^{*} \Phi \biggr],
\label{AH5c}
\end{eqnarray}
where $\Pi_{1}$ and $\Pi_{2}$ are defined, respectively, as $\Pi_{1} = \Phi^{*\, \prime} - {\mathcal H} ( 1 + 6 \xi) \Phi^{*}$  and as $\Pi_{2} = \Phi^{*\, \prime}$.
The Hamilton equations derived either from Eq. (\ref{AH5b}) or from Eq. (\ref{AH5c}) are the same 
since the Hamiltonians are related by a canonical transformation. In the  limit
$\tau \to -\infty$, Eqs. (\ref{AH5b}) and (\ref{AH5c}) coincide, since, in this limit,  $\Pi_{1} \sim \Pi_{2}$. 
If  quantum-mechanical initial conditions are assigned for 
$\tau \to -\infty$,  the state minimizing $H_{1}$ will also minimize $H_{2}$. 
On the contrary, when initial conditions are imposed at   
a finite value of the conformal time coordinate  the states minimizing $H_{1}$ and $H_{2}$ 
might differ \cite{assi}. These differences are immaterial for the problem at hand so that we shall consistently carry on the 
quantization (and the evaluation of the expectation values) in terms of $H_{2}$.

\subsection{Quantum Hamiltonian}
Promoting the classical fields to quantum operators and representing the operators in Fourier space we can then write 
\begin{equation}
\hat{\Phi}(\vec{x},\tau) = \int \frac{d^{3} k}{(2\pi)^{3/2}} \,\hat{\Phi}_{\vec{k}}(\tau)\, e^{- i \vec{k}\cdot\vec{x}}, \qquad 
\hat{\Pi}(\vec{x},\tau) = \int \frac{d^{3} k}{(2\pi)^{3/2}} \,\hat{\Pi}_{\vec{k}}(\tau)\, e^{- i \vec{k}\cdot\vec{x}},
\label{AH6}
\end{equation}
so that,  from Eq. (\ref{AH5c}), the Hamiltonian relevant to our problem becomes
\begin{eqnarray}
\hat{H}(\tau) =  \int d^{3} k \biggl[ \hat{\Pi}_{\vec{k}}^{\dagger} \hat{\Pi}_{\vec{k}} + 
\omega_{k}^2 \hat{\Phi}_{\vec{k}}^{\dagger} \hat{\Phi}_{\vec{k}} 
- (6 \xi + 1) ({\mathcal H}^2 + {\mathcal H}^{\prime}) \hat{\Phi}_{\vec{k}}^{\dagger} \hat{\Phi}_{\vec{k}} \biggr],
\label{AH7} 
\end{eqnarray}
where $\omega_{k}^2(\tau) = k^2 + m^2 a^2(\tau)$. For each mode of the field, Eq. (\ref{AH7}) provides the quantum description 
of the process of parametric amplification originally analyzed in the context of quantum optics \cite{mollow}.
The evolution equations in the Heisenberg description can be easily obtained from Eq. (\ref{AH7}):
\begin{eqnarray}
\partial_{\tau}\hat{\Phi}_{\vec{k}} = \hat{\Pi}^{\dagger}_{\vec{k}}, \qquad 
\partial_{\tau}\hat{\Pi}_{\vec{k}}^{\dagger} = - \omega_{k}^2 \hat{\Phi}_{\vec{k}} + (1+ 6 \xi) \frac{a^{\prime\prime}}{a} \hat{\Phi}_{\vec{k}},
\label{AH9}
\end{eqnarray}
and similarly for the Hermitian conjugate operators. The field operators $\hat{\Phi}_{\vec{k}}$ and $\hat{\Pi}_{\vec{k}}$ can 
be expressed in terms of the creation and annihilation operators $\hat{a}_{\vec{k}}(\tau)$ and $\hat{b}_{\vec{k}}(\tau)$, namely
\begin{eqnarray}
\hat{\Phi}_{\vec{k}} &=& \frac{1}{\sqrt{2 \omega_{k}}} ( \hat{a}_{\vec{k}} + \hat{b}_{-\vec{k}}^{\dagger}),\qquad \hat{\Pi}_{\vec{k}} = - i \sqrt{\frac{\omega_{k}}{2}} ( \hat{b}_{\vec{k}} - \hat{a}_{-\vec{k}}^{\dagger}).
\label{AA3}
\end{eqnarray}
Where $\hat{a}_{\vec{k}}$ and $\hat{b}_{\vec{p}}$ and separately obey the standard commutation relations, namely, $[\hat{a}_{\vec{k}}, \, \hat{a}^{\dagger}_{\vec{p}}] =
\delta^{(3)}(\vec{k} - \vec{p})$ and $[\hat{b}_{\vec{k}}, \, \hat{b}^{\dagger}_{\vec{p}}] = \delta^{(3)}(\vec{k} - \vec{p})$. The Hermitian conjugate relations can be directly derived from Eq. (\ref{AA3}).

The quantum Hamiltonian of Eq. (\ref{AH7}) could now be expressed in terms of the 
creation and annihilation operators $\hat{a}_{\vec{k}}$ and $\hat{b}_{\vec{k}}$;
the result is not diagonal but it can be brought to diagonal form by means 
of the following canonical transformation:
\begin{eqnarray}
\hat{a}_{\vec{k}}(\tau) &=& u_{k}(\tau) \, \hat{a}_{\vec{k}}(\tau_{i}) - v_{k}(\tau)\,  \hat{b}^{\dagger}_{-\vec{k}}(\tau_{i}),
\label{AA5}\\
\hat{b}_{-\vec{k}}(\tau) &=& u_{k}(\tau) \, \hat{b}_{-\vec{k}}(\tau_{i}) - v_{k}(\tau)\,  \hat{a}^{\dagger}_{\vec{k}}(\tau_{i}),
\label{AA6}
\end{eqnarray}
where $|u_{k}(\tau)|^2 - |v_{k}(\tau)|^2 =1$.  The same canonical transformation 
is used to derive the ground state wavefunction of an interacting Bose gas at zero 
temperature \cite{mb}. In terms of the  operators  $\hat{a}_{\vec{k}}(\tau_{i})$ and $\hat{b}_{\vec{k}}(\tau_{i})$ the Hamiltonian is diagonal.

When the modes of the field are inside the Hubble radius at $\tau_{i}$ (i.e. $k\tau_{i} \gg 1$) the vacuum corresponds 
 to the state  $| s_{i} \rangle = |0_{a}\,\, 0_{b} \rangle$ which is annihilated both by $\hat{a}_{\vec{k}}(\tau_{i})$ and 
 by $\hat{b}_{\vec{k}}(\tau_{i})$. From Eqs. (\ref{AA5}) and (\ref{AA6}) the average multiplicity of produced Higgs excitations per Fourier mode 
 at a generic time $|\tau| \gg \tau_{i}$ is given by:
 \begin{equation}
 \overline{n}_{k}(\tau) = \langle s_{i} | \hat{a}_{\vec{k}}^{\dagger}(\tau) \hat{a}_{\vec{k}}(\tau) + \hat{b}_{\vec{k}}^{\dagger}(\tau) \hat{b}_{\vec{k}}(\tau) | s_{i} \rangle = 2 |v_{k}(\tau)|^2.
\label{AA6a}
\end{equation}
Equation (\ref{AA6a}) accounts of the number of produced excitations; the particle 
content of the initial state will be disregarded  since the average multiplicity of the produced quanta is typically very large the contribution of the initial state 
is immaterial for the present ends; it may be relevant to gauge the backreaction of the initial state as suggested in \cite{assi}.

The evolution of $u_{k}(\tau)$ and $v_{k}(\tau)$ is related to the mode functions for the canonical fields and for the canonical momenta, i.e. 
$u_{k} - v_{k}^{*} = \sqrt{2 \omega_{k}} F_{k}$ and $u_{k} + v_{k}^{*} = i G_{k}\sqrt{2/\omega_{k}}$ where 
$F_{k}$ and $G_{k}$ obey:
\begin{equation}
F_{k}^{\prime} = G_{k} , \qquad G_{k}^{\prime} = - \omega_{k}^2 F_{k} + (1 + 6 \xi) ({\mathcal H}^2 + {\mathcal H}^{\prime}) F_{k}.
\label{AA7}
\end{equation}
Denoting with $\rho_{\Phi}$ 
the energy density of the produced inhomogeneities we also have
$d \rho_{\Phi} = \omega_{k} \overline{n}_{k} d^{3} k/(2 \pi)^3$ which is valid in the limit $\overline{n}_{k} \gg 1$.When the background geometry expands
from the inflationary stage to the radiation epoch the effective Higgs mass gives a negligible contribution 
during inflation. In the radiation phase the contribution of $\xi$ can be totally disregarded (since the Ricci scalar 
vanishes on the background) but the mass contribution increases as $m^2 a^2$ and eventually dominates the evolution 
of the mode functions.

\renewcommand{\theequation}{3.\arabic{equation}}
\setcounter{equation}{0}
\section{Production of massive modes and their current} 
\label{sec3}
If he slow-roll parameters 
are constant during the quasi-de Sitter phase the connection between the  
conformal time coordinate and  the Hubble rate is simpler, namely ${\mathcal H} = aH = -1/[(1-\epsilon)\tau]$. 
The inflationary scale factor can then be expressed as 
$a_{i}(\tau) = (- \tau/\tau_{1})^{- \beta}$ for $\tau < - \tau_1$, where $-\tau_{1}$ 
marks the end of the inflationary phase and $\beta=1$ in the case a de Sitter phase. 
The scale factor of the radiation epoch hold for  $\tau\geq -\tau_{1}$ and it is given by
$a_{r}(\tau) = [\beta \tau + (\beta + 1)\tau_1]/\tau_1$. 
The scale factors and their first derivatives with respect to $\tau$ are continuously matched across the transition, 
[i.e.  $a_{r}(-\tau_{1}) = a_{i}(-\tau_{1})$ and $a^{\prime}_{r}(-\tau_{1}) = a^{\prime}_{i}(- \tau_{1})$] so that 
the extrinsic curvature is also continuous [i.e. ${\mathcal H}_{r}(-\tau_{1}) = {\mathcal H}_{i}(- \tau_{1})$].

Instead of constructing a continuous background by direct matching it is possible to find 
an interpolating background joining the two regions. This was the strategy in various applications 
of quantum field theory in curved backgrounds \cite{birrel} and the same strategy has been 
used recently in a similar context \cite{rec}. In the present case the interpolating background could be, for instance 
\begin{equation}
a(\tau) = a_{1}( \tau + \sqrt{\tau^2 + \tau_{1}^2}).
\label{int}
\end{equation}
For $\tau \ll -\tau_{1}$ the scale factor goes as $a(\tau) \simeq (-\tau_{1}/\tau)$ (quasi de-Sitter expansion); 
conversely, if   $\tau \gg \tau_{1}$, $a(\tau) \simeq (\tau/\tau_{1})$ (radiation dominated evolution). 
It is clear from the analytic form of Eq. (\ref{int}) that for large negative times (i.e. $\tau \ll - \tau_{1}$)
$a(\tau) \simeq (- \tau_{1}/\tau)$. Conversely for $\tau \gg \tau_{1}$ we will have that $a(\tau) \simeq (\tau/\tau_1)$.
This means that this background has the correct asymptotic behaviour and it is compatible with a time dependent 
equation of state. Now this strategy is clearly equivalent (but probably less general) than the one adopted here where 
the inflationary phase is genuinely described in terms of the slow-roll parameters. The inhomogeneities 
of the Higgs field will clearly be amplified also in the case of Eq. (\ref{int}) since what matters is that 
the expansion rate changes between two asymptotic regions of the evolution of the background. 

\subsection{Non-minimal coupling in the mode functions}
Using the explicit form of the scale factor during inflation given in the previous paragraph, 
the solution of Eq. (\ref{AA7}) implies that $F_{k}(\tau)$
and $G_{k}(\tau)$ for $\tau< -\tau_{1}$ are given by:
\begin{eqnarray}
F_{k}(\tau) &=& f_{k}(\tau) = \frac{{\mathcal N}}{\sqrt{2 k}} \sqrt{- k \tau } H^{(1)}_{\alpha}(- k\tau),\qquad \tau < - \tau_{1}
\label{BB1}\\
G_{k}(\tau) &=& g_{k}(\tau) ={\mathcal N}\sqrt{\frac{k}{2}} \sqrt{-k\tau}\biggl[ H^{(1)}_{\alpha -1} (- k \tau) +
\frac{(1 -2 \alpha)}{2(- k \tau)} H^{(1)}_{\alpha} (- k \tau)\biggr], \qquad \tau < - \tau_{1},
\label{BB2}
\end{eqnarray}
where $H_{\alpha}^{(1)}(-k\tau)$ is the Hankel function of first kind \cite{abr} with index $\alpha$ and argument $-k\tau$.
In explicit terms $\alpha$ and ${\mathcal N}$ are defined as: 
\begin{equation}
\alpha =\frac{1}{2} \sqrt{ 1 +  \frac{4( 2 - \epsilon) ( 1 + 6 \xi)}{(1 - \epsilon)^2} - \frac{\zeta^2}{(1 - \epsilon)^2} },\qquad {\mathcal N}= \sqrt{\frac{\pi}{2}}\,\,e^{i\frac{\pi}{4}( 1 + 2 \alpha)}, 
\label{BB3}
\end{equation}
where $\zeta = m/(a_{e} H_{e}) \ll 1$ is the (constant) ratio between the mass and the inflationary expansion rate while $\epsilon$ is the standard slow-roll
parameter already introduced after Eq. (\ref{sc2}). To lowest order in $\epsilon$, the second relation in Eq. (\ref{BB3}) implies:
\begin{equation}
\alpha = \frac{3}{2} \sqrt{1 + \frac{16}{3} \xi - \frac{\zeta^2}{9} } + \frac{(6 - \zeta^2 + 36 \xi)\epsilon}{2 \sqrt{9 - \zeta^2 + 48 \xi}} + {\mathcal O}(\epsilon^2),
\label{BB3a}
\end{equation}
with the proviso that the condition $\zeta \ll 1$ is always verified in practice. For the sake of simplicity we shall require $\xi > -3/16$ so that the expansion (\ref{BB3a}) will always be well defined when $\zeta$ is negligible during inflation.

\subsection{Massive and relativistic modes}
When all the modes are ultrarelativistic (i.e. $k \gg m a$) the solutions of Eq. (\ref{AA7}) in the radiation epoch are simply plane waves.
Moreover, since $a^{\prime\prime}=0$  the mode functions are not affected by the presence of  $\xi$. Over large-scales the contribution of the 
massive modes is comparatively more relevant. The same separation of scales occurs in the minimally coupled limit where $\xi \to 0$ \cite{mg2}.
The mode functions  for $\tau \geq -\tau_{1}$ can be expressed as: 
\begin{eqnarray}
F_{k}(\tau) &=& \mu_{k} \,\,\widetilde{f}_{k}(\tau) + \nu_{k} \,\,\widetilde{f}_{k}^{*}(\tau),\qquad \tau \geq - \tau_{1},
\label{BB4}\\
G_{k}(\tau) &=& \mu_{k} \,\,\widetilde{g}_{k}(\tau) + \nu_{k} \,\,\widetilde{g}_{k}^{*}(\tau),\qquad \tau \geq - \tau_{1},
\label{BB5}
\end{eqnarray}
where $\widetilde{f}_{k}(\tau)$ and  $\widetilde{g}_{k}(\tau)$ are solutions of Eq. (\ref{AA7}) in the 
radiation-dominated epoch when the scale factor is given by $a_{r}(\tau) = [\beta \tau + (\beta + 1)\tau_1]/\tau_1$; the 
explicit expressions of  $\widetilde{f}_{k}(\tau)$ and  $\widetilde{g}_{k}(\tau)$ are:
\begin{eqnarray}
\widetilde{f}_{k}(\tau) &=& \frac{1}{\sqrt[4]{2 \gamma} } \, e^{i\frac{\pi}{8}}  {\mathcal D}_{\sigma}( e^{i\frac{\pi}{4}} z), \qquad \sigma=  - (i p + 1/2),
\nonumber\\
 \widetilde{g}_{k}(\tau) &=& \sqrt[4]{2\gamma} e^{3i\frac{\pi}{8}} \biggl[ \frac{z}{2}\,  e^{i\frac{\pi}{4}} \, {\mathcal D}_{\sigma}(e^{i\frac{\pi}{4}} z) - {\mathcal D}_{\sigma+1}(e^{i\frac{\pi}{4}} z) \biggr],
\label{BB6}
\end{eqnarray}
where ${\mathcal D}_{\sigma}(x)$ are the parabolic cylinder functions with index $\sigma$ and argument $x$ \cite{abr}. Note that  $\widetilde{f}_{k}^{*}(z)= {\mathcal D}_{i p - 1/2}( e^{3 i \pi/4} z)/\sqrt[4]{2\gamma}$; the variables $z$, $p$ and $\gamma$ appearing in Eq. (\ref{BB6}) are defined as: 
\begin{equation}
z = \sqrt{2 \gamma} \biggl[ \tau + \frac{(\beta+1)}{\beta}\tau_1\biggr],\qquad p = \frac{k^2}{2\gamma}, \qquad \gamma=\frac{\zeta \beta}{\tau_{1}^2}.
\label{BB7}
\end{equation}
The phases have been selected in such a way that for $z \gg | p|$ and for $k^2 \tau_1 \ll m$, 
$\widetilde{f}_{k}(\tau) \to \sqrt{ \tau_1/(2 m \tau)} \exp{[ - i m \tau^2/(2 \tau_1)]}$. 
In terms of these variables the evolution equation for $\widetilde{f}_{k}$ 
can be written, from Eq. (\ref{AA7}) as $ d^2\widetilde{f}_{k}/d z^2 + [ p + z^2/4] \widetilde{f}_{k} =0$.
\subsection{Mixing coefficients}
The  continuous and differentiable transition\footnote{The pump field contains second derivatives of the scale factor. The lack of continuity of the scale factor and of its first (conformal) time derivative would imply a singularity in $a^{\prime\prime}$. If $a$ and $a^{\prime}$ are both continuous
$a^{\prime\prime}$ contains, at most, a discontinuity.} between the inflationary phase and the radiation dominated epoch implies that
 the expressions of the mode functions given in Eqs. (\ref{BB1})--(\ref{BB2}) and in Eqs. (\ref{BB3})--(\ref{BB4}) must also be 
continuous in $\tau= - \tau_{1}$. Consequently the continuity of $F_{k}(\tau)$ and $G_{k}(\tau)$ implies
\footnote{Note that the Wronskian of the solution must be conserved in the two regimes (i.e. 
$F_{k} G_{k}^{*} - F_{k}^{*} G_{k} = i$), we have that $|\mu_{k}|^2 - |\nu_{k}|^2 =1$.}:
\begin{eqnarray}
f_{k}(-\tau_{1}) &=& \mu_{k} \widetilde{f}_{k}(-\tau_{1}) + \nu_{k} \widetilde{f}_{k}^{*}(-\tau_{1}),
\nonumber\\
g_{k}(-\tau_{1}) &=& \mu_{k} \widetilde{g}_{k}(-\tau_{1}) + \nu_{k} \widetilde{g}_{k}^{*}(-\tau_{1}).
\label{BB8}
\end{eqnarray}
By solving this system the expression  of $\mu_{k}$ and $\nu_{k}$ are functions of two dimensionless 
variables $m\,\tau_1$ (coinciding with $\zeta$ since $\tau_{1} \equiv \tau_{e}$) and $k\,\tau_{1}$. For the sake of accuracy  
it is appropriate to expand $\mu_{k}$ and $\nu_{k}$  in the limit $\zeta \ll 1$ (for fixed $k\tau_{1}$). The result of this 
manipulation is:
\begin{eqnarray}
\mu_{k} &=& \frac{ e^{ i \pi(\alpha + 1/4)/2}  \Gamma(1/4)}{ 4 \,\beta^{1/4}\, \zeta^{1/4}}  \biggl[ \biggl( \alpha + \frac{1}{2}\biggr) 
H_{\alpha}^{(1)}(k\tau_1) - k\tau_1 H^{(1)}_{\alpha + 1 } (k\tau_1)\biggr]
\nonumber\\
&-& i \frac{(1 + i)e^{ i \pi(\alpha + 1/4)/2}\Gamma(3/4) }{ 2 \sqrt{2} \beta^{3/4}}  \biggl[ \biggl( \alpha +\beta + \frac{1}{2}\biggr) 
H_{\alpha}^{(1)}(k\tau_1) - k \tau_1 H^{(1)}_{\alpha + 1 } (k\tau_1)\biggr] \zeta^{1/4} + {\mathcal O}(\zeta^{5/4}),
\nonumber\\
\nu_{k} &=&  - \frac{ e^{ i \pi(\alpha + 3/4)/2}\Gamma(1/4)}{ 4 \,\beta^{1/4}\, \zeta^{1/4} }  \biggl[ \biggl( \alpha + \frac{1}{2}\biggr) 
H_{\alpha}^{(1)}(k\tau_1) - k\tau_1 H^{(1)}_{\alpha + 1 } (k\tau_1)\biggr]
\nonumber\\
&+&\frac{(1 + i)e^{ i \pi(\alpha + 3/4)/2}  \Gamma(3/4)}{ 2 \sqrt{2} \beta^{3/4}} \biggl[ \biggl( \alpha +\beta + \frac{1}{2}\biggr) 
H_{\alpha}^{(1)}(k\tau_1)
\nonumber\\
&-& k \tau_1 H^{(1)}_{\alpha + 1 } (k\tau_1)\biggr] \zeta^{1/4} + {\mathcal O}(\zeta^{5/4}).
\label{nu}
\end{eqnarray}
Equation (\ref{nu}) can be further expanded in the limit $| k \tau_{1}| \ll 1$  by recalling the small argument limit of the 
corresponding Hankel functions \cite{abr}.  When the modes are ultrarelativistic 
mode functions are plane waves and the mixing coefficients have a well known 
form\footnote{The mode functions $\widetilde{f}_{k}(\tau)$ are plane waves with argument $k [\tau + (\beta + 1)\tau_{1}/\beta]$. 
The dependence on $\xi$ drops also in the relativistic branch of the spectrum during the radiation epoch. The expression of the coefficients 
$\mu_{k}$ and $\nu_{k}$ are modified and in the case $\alpha \to 3/2$ they are given by
$\mu_{k} = e^{2 i k\tau_{1}} [ 1 - i/(k\tau_{1})- 1/(2 k^2 \tau_{1}^2)]$ and $\nu_{k} = 1/(2 k^2 \tau_{1}^2)$.}.

\subsection{Currents and average multiplicities}
With the explicit expressions of the mode functions and of the mixing coefficients the charge and the current density fluctuations can be directly computed from Eq. (\ref{AH4a}). The result of this calculation can be expressed in rather simple terms:
\begin{eqnarray}
\hat{\rho}(\vec{x},\tau) &=& \frac{i q}{ (2 \pi)^3} \int d^{3} k \int d^{3} p \biggl[ \hat{\Phi}^{\dagger}_{-\vec{k}}(\tau)\, \hat{\Pi}^{\dagger}_{- \vec{p}}(\tau)  - \hat{\Phi}_{\vec{k}}(\tau)\,\hat{\Pi}_{\vec{p}}(\tau) \biggr],
\label{CC1}\\
\hat{J}_{i}(\vec{x}, \tau) &=& \frac{q}{(2 \pi)^3} \int d^{3} k \int d^{3} p (p_{i} - k_{i}) \hat{\Phi}_{\vec{k}}(\tau) \hat{\Phi}^{\dagger}_{-\vec{p}}(\tau).
\label{CC2} 
\end{eqnarray}
From Eq. (\ref{CC2}) we can also easily deduce that
\begin{equation}
(\vec{\nabla}\times \vec{J})_{k} = - \frac{i q\,\,\epsilon_{i j k}}{(2 \pi)^3} \int d^{3} k \int d^{3} p \, ( k_{i} + p_{i}) \, ( p_{j} - k_{j}) \, \hat{\Phi}_{\vec{k}}(\tau) \hat{\Phi}_{-\vec{p}}^{\dagger}(\tau) \, \, e^{ - i (\vec{k} + \vec{p})\cdot \vec{x}}.
\label{CC3}
\end{equation}
Using Eqs. (\ref{CC1}), (\ref{CC2}) and (\ref{CC3}) the corresponding correlation functions can be computed. They all involve the expectation values of 
four field operators. For the present purposes we shall be particularly interested in the following correlation function\footnote{ We note that the current 
induced by the homogeneous mode of the Higgs vanishes; this is obvious from Eqs. (\ref{CC4}) and (\ref{ChR}). The corresponding magnetic field will
also vanish \cite{mg2}.}
\begin{eqnarray}
\langle (\vec{\nabla}\times \vec{J})_{(\vec{x},\tau)} \cdot (\vec{\nabla}\times \vec{J})_{(\vec{y},\tau)}\rangle  &=& \frac{q^2}{(2\pi)^{6}}\int \frac{d^{3} k}{\omega_{k}} \int 
\frac{d^{3} k^{\,\prime}}{\omega_{k^{\,\prime}}} [ k^2 k^{\,\prime\, 2}- (\vec{k}\cdot\vec{k}^{\,\prime})^2] {\mathcal C} (k, k^{\prime}, \tau)
 e^{- i (\vec{k} + \vec{k}^{\prime})\cdot\vec{r}},
\nonumber\\
{\mathcal C}(k, k^{\prime}, \tau) &=& |u_{k}(\tau)- v_{k}^{*}(\tau)|^2 \,  |u_{k^{\prime}}(\tau)- v_{k^{\prime}}^{*}(\tau)|^2.
\label{CC4}
\end{eqnarray}
Equation (\ref{CC4}) and its descendants are derived by using the explicit expressions of the field operators and by evaluating 
the corresponding expectation values. As remarked in Eqs. (\ref{AA5}) and (\ref{AA6}) 
 the initial state $|\,s_{i} \rangle$ is annihilated by $\hat{a}_{\vec{k}}(\tau_{i})$ and $\hat{b}_{\vec{k}}(\tau_{i})$ 
and not simply by $\hat{a}_{\vec{k}}(\tau)$ and $\hat{b}_{\vec{k}}(\tau)$.

The problem of the regularization of these correlators has been discussed in detail in Ref. \cite{mg2}
and we shall use here the same computational scheme. Defining the  
Gaussian window function $W(\vec{x},\tau) = \exp{- [k_{L}^2 |\vec{x}|^2/2 + \omega_{L}^2 \tau^2/2]}$
 the regularized charge and current density fluctuations will be given by:
\begin{eqnarray}
&& (\vec{\nabla}\times\vec{{\mathcal J}})^2 =  \Omega_{L}^2 k_{L}^6 \int d^3x \int d^3y \int d\tau \int d\tau^{\prime} 
\langle(\vec{\nabla} \times \vec{J})\cdot(\vec{\nabla}\times \vec{J})\rangle W(\vec{x},\Delta\tau) W(\vec{y},\Delta\tau^{\prime}),
\nonumber\\
&& {\mathcal Q}^2 = \Omega_{L}^2 k_{L}^6 \int d^3 x \int d^3 y\int d\tau \int d\tau^{\prime} \langle 
\hat{\rho}( \vec{x}, \tau) \hat{\rho} (\vec{y}, \tau^{\prime}) 
\rangle W(\vec{x},\Delta\tau) W(\vec{y},\Delta\tau^{\prime}),
\label{ChR}
\end{eqnarray}
where $\Delta\tau =(\tau-\tau_{R})$ and $\Delta\tau^{\prime} =(\tau^{\prime} - \tau_{R})$; the smearing functions select the
contribution of the  correlators inside a given space-time region with typical comoving volume $V \simeq 1/k_{L}^3$ and 
over a comoving time $1/\Omega_{L}$; $V$ and $\Omega_{L}$ will be chosen in such a way that 
$\Omega_{L}^{-3} \simeq V \ll m^{-3}$; furthermore we shall choose $\tau_{R} \geq |\tau_1|$ where $\tau_{1}$ denotes 
the inflationary boundary. 

We are now going to estimate the average multiplicity of the Higgs inhomogeneities.
The average multiplicity determines the energy density of the produced 
Higgs quanta (denoted by $\rho_{\Phi}$ hereunder) and  we have to make sure that the energy density 
of the Higgs quanta does not exceed the energy density 
of the background. If this would happen the Higgs would 
cease to be a spectator field; this is not the case and $\rho_{\Phi}$ 
 is always much smaller than the background contribution. 
Indeed, the average multiplicity of the produced Higgs particles can be computed from Eq. (\ref{AA6a}) by taking into account the asymptotic 
form of the mode functions (see Eq. (\ref{BB6})) and the results of Eq. (\ref{nu}). The 
final expression for the average multiplicity and for the energy density  can be expressed as\footnote{Since the end of inflation coincides with the onset 
of the radiation epoch we have that, by definition, $|k\tau_{1}| = | k/(a_{e} H_{e})|$.} :
\begin{equation}
\overline{n}_{k} \simeq  \delta \,\biggl(\frac{k}{a_{e} H_{e}}\biggr)^{-2 \alpha} \zeta^{-1/2}, \qquad \frac{d\rho_{\Phi}}{d \ln{k}} \simeq \delta \,m\, H_{e}^3\, \biggl(\frac{k}{a_{e} H_{e}}\biggr)^{3 - 2\alpha} \zeta^{-1/2} \biggl(\frac{a_{e}}{a}\biggr)^3
\label{expl1}
\end{equation}
where $\delta = {\mathcal O}(10^{-2})$.  In  the case of quasi-flat spectrum (i.e.  $\alpha\simeq 3/2$) up to an irrelevant logarithmic correction the energy density of the massive quanta scales as $\rho_{\Phi}(\tau) = \delta\, m \,H_{e}^3 (a_{1}/a)^3$. Because the energy density of the radiation background redshifts as 
$\rho_{r}(\tau) \sim H_{e}^2 M_{P}^2 (a_{1}/a)^4$ we have that 
$\rho_{\Phi}$ will eventually exceed $\rho_{r}$ at a given time $\tau_{2}$ where $a_{2}/a_{1} \simeq (H_{e}/M_{P})^2 \zeta$. For instance, in the case 
$m = m_{0}= {\mathcal O}(125)$ GeV the corresponding expansion rate would be $H_{2} \simeq {\mathcal O}(10^{-32}) H_{e}$. 
This potential constraint is however invalidated since the Higgs 
condensate decays\footnote{The Standard Model Higgs could decay perturbatively into quarks, leptons and gauge fields. The decay channels into weak gauge bosons and top quarks are kinematically blocked.  Since the decay rate is proportional to the square of the fermion mass, the decay into bottom quarks dominate. The non-perturbative decay is much more efficient \cite{k1} and faster than its perturbative counterpart.} before $\tau_{2}$. The Higgs gets effectively massive at a scale $m \simeq H_{os}$: this equation should be viewed as a definition of $H_{os}$. Note that $m$ (as already mentioned 
after Eq. (\ref{AH3})) is the effective Higgs mass. Using the results of Ref. \cite{k1} we have that $H_{os} \simeq 0.25 \lambda^{3/4} H_{e}$. 
For $\lambda\simeq 0.01$ this result implies that $H_{os} \simeq 
8 \times 10^{-3} H_{e}$ and that $H_{os}/H_{dec} = {\mathcal O}(300)$ where $H_{dec}$ denotes the expansion 
rate at decay (see also \cite{k1}). Consequently, since $H_{dec} \simeq 2.6 \times 10^{-5} H_{e}$ we have that $H_{2} \gg H_{dec}$, as anticipated.

The considerations developed so far hold for modes $m a > k$ during the radiation epoch.  The massless modes are only relevant over short wavelengths and 
their average multiplicity will be given, approximately, by $\overline{n}_{k} \simeq \delta |k/(a_{e} H_{e})|^{-2\alpha}$. Recalling that 
$ d n = \overline{n}_{k} d^{3}k/{(2\pi)^3}$ we have that the total concentration (for simplicity in the case $\alpha\simeq 3/2$) will 
be, approximately, $ n = \delta H_{e}^3 (a_{e}/a)^3$ while their energy density will be, as usual, $\rho_{\Phi} \simeq \delta \,H_{e}^4 (a_{e}/a)^{4}$. 

\renewcommand{\theequation}{4.\arabic{equation}}
\setcounter{equation}{0}
\section{Bounds on the non-minimal coupling} 
\label{sec4}
The evolution of the gauge fields across the radiation phase is customarily analyzed by using the standard kinetic description 
of the plasma \cite{pl1} appropriately generalized to conformally flat backgrounds 
 \cite{pl2}. While the magnetohydrodynamical treatment is only meaningful for sufficiently 
 low frequencies,  the relaxation of an initial inhomogeneities is a high-frequency phenomenon  involving scales 
 close to the plasma frequency.  The regularized charge and current density fluctuations of Eq. (\ref{ChR}) 
serve as the initial conditions of the Einstein-Vlasov system during the post-inflationary evolution. 
In this framework the charge fluctuations and the corresponding electric fields are efficiently dissipated but the magnetic fields are not suppressed for 
wavenumbers smaller than the magnetic diffusivity scale $k_{\sigma} \simeq \sqrt{\sigma_{c} {\mathcal H}}$ where $\sigma_{c}$ denotes the conductivity of the post-inflationary plasma. For $k < k_{\sigma}$ the Einstein-Vlasov system implies \cite{mg2} that the comoving magnetic energy density relaxes to\footnote{The non-screened vector modes of the hypercharge
field project on the electromagnetic fields through the cosine of the Weinberg angle (denoted by $\cos{\theta_{W}}$ in Eq. (\ref{BB})). Note that the present model of magnetic field generation has nothing to do with models where the inflaton directly couples with the 
kinetic term of the gauge fields \cite{DT1,DT2,DT3}. It has been recently shown that these models can be generalized to the case where the electric and  magnetic gauge couplings have different evolutions \cite{VdW}, as it happens in the relativistic theory of Van der Waals interactions; viable scenarios free of strong coupling problems can be constructed in this framework.} 
 \begin{equation}
 \rho_{B} \to \frac{T^2}{32\pi^2 \,\overline{\alpha}^2 \,n_{c}^2} \biggl(\vec{\nabla}\times \vec{{\mathcal J}}\biggr)^2\,\cos^2{\theta_{W}},
 \label{BB}
 \end{equation}
 where, as already mentioned, the currents have been regularized according to Eq. (\ref{ChR}); $n_{c}(T)$ is concentration of the charged particles with average momentum $T$ and $\overline{\alpha}$ is the gauge coupling constant \cite{mg2} not to be confused with the Hankel index $\alpha$ which depends on $\epsilon$, $\zeta$ and $\xi$ (see Eq.  (\ref{BB3}) and discussion thereafter). It should be mentioned 
 that, in the present treatment, the same information contained in the energy density enters the power spectrum of the magnetic field.
    
A detailed description  of the sudden rise of the conductivity across the inflationary transition may depend on the dynamics of reheating, on the couplings of the inflaton and even on the protoinflationary initial conditions (see e.g. \cite{mg0}). The rate of equilibration of the electromagnetic reactions is however large in comparison with the expansion rate after inflation (and may be even bigger than the rate of the inflaton decay). Since the obtained results depend solely on the density of the charged particles and on their typical momentum, Eq. (\ref{BB}) is expected to be qualitatively applicable also when different species are not in local thermal equilibrium but only in kinetic equilibrium at different temperatures \cite{mg2}.
In the simplest physical situation the protoinflationary initial data do not contain any globally neutral plasma and the number of efolds is minimal or close to minimal (see Eq. (\ref{xx}) and discussion thereafter).  A simple model of the conducting transition can be found, for instance, in \cite{mg0} (see also Eq. (\ref{int})):
\begin{equation}
a(\tau) = a_{1} ( \tau + \sqrt{\tau^2 + \tau_1}),\qquad \sigma_{\mathrm{c}}(\tau) = \frac{T}{\overline{\alpha}} \theta(\tau),\qquad
 \theta(\tau) = \frac{1}{8} \biggl( 1 + \frac{\tau}{\sqrt{\tau^2 + \tau_{1}^2}}\biggr)^{3}.
\label{MM1}
\end{equation}
Note that $\theta(\tau)$ is a smooth representation of the Heaviside step function and $\sigma_{\mathrm{c}}(\tau)$ is the conductivity which is of the order of $ T/\overline{\alpha}$ where $T$ is the temperature at the corresponding 
epoch. The sudden transition is recovered in the formal limit $|\tau_{1}| \to 0$. 
\begin{figure}[!ht]
\centering
\includegraphics[height=7cm]{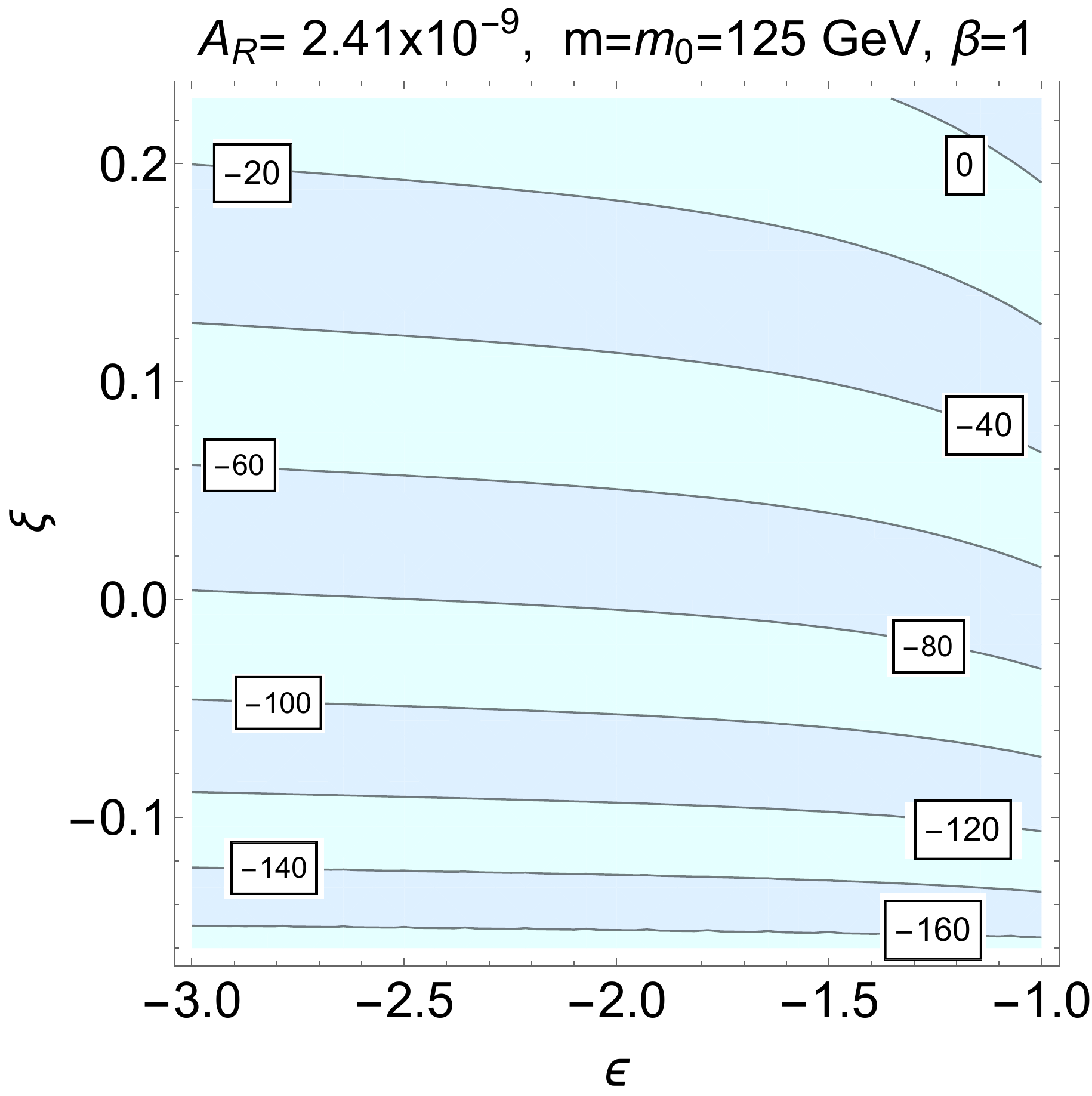}
\includegraphics[height=7cm]{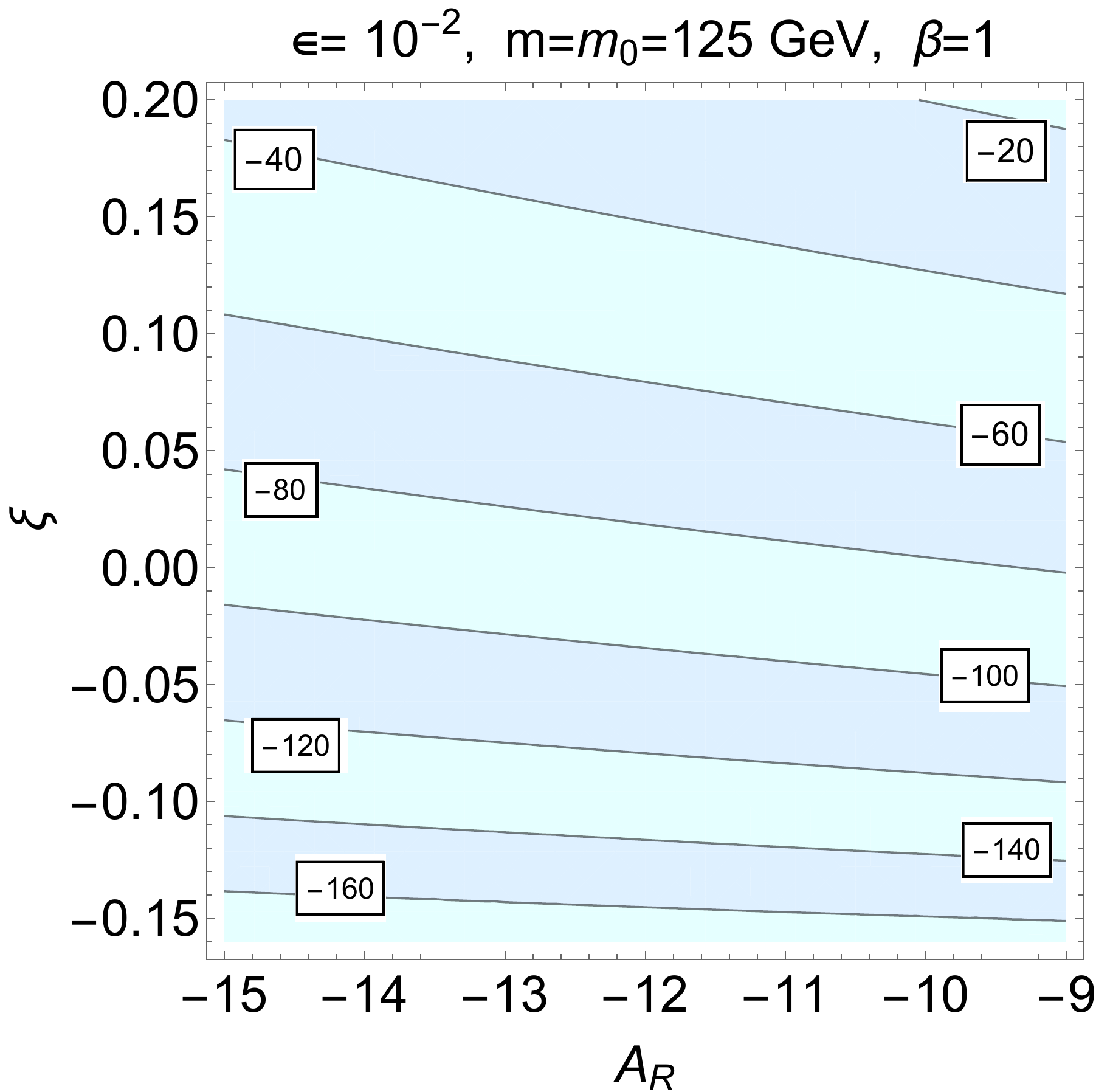}
\includegraphics[height=7cm]{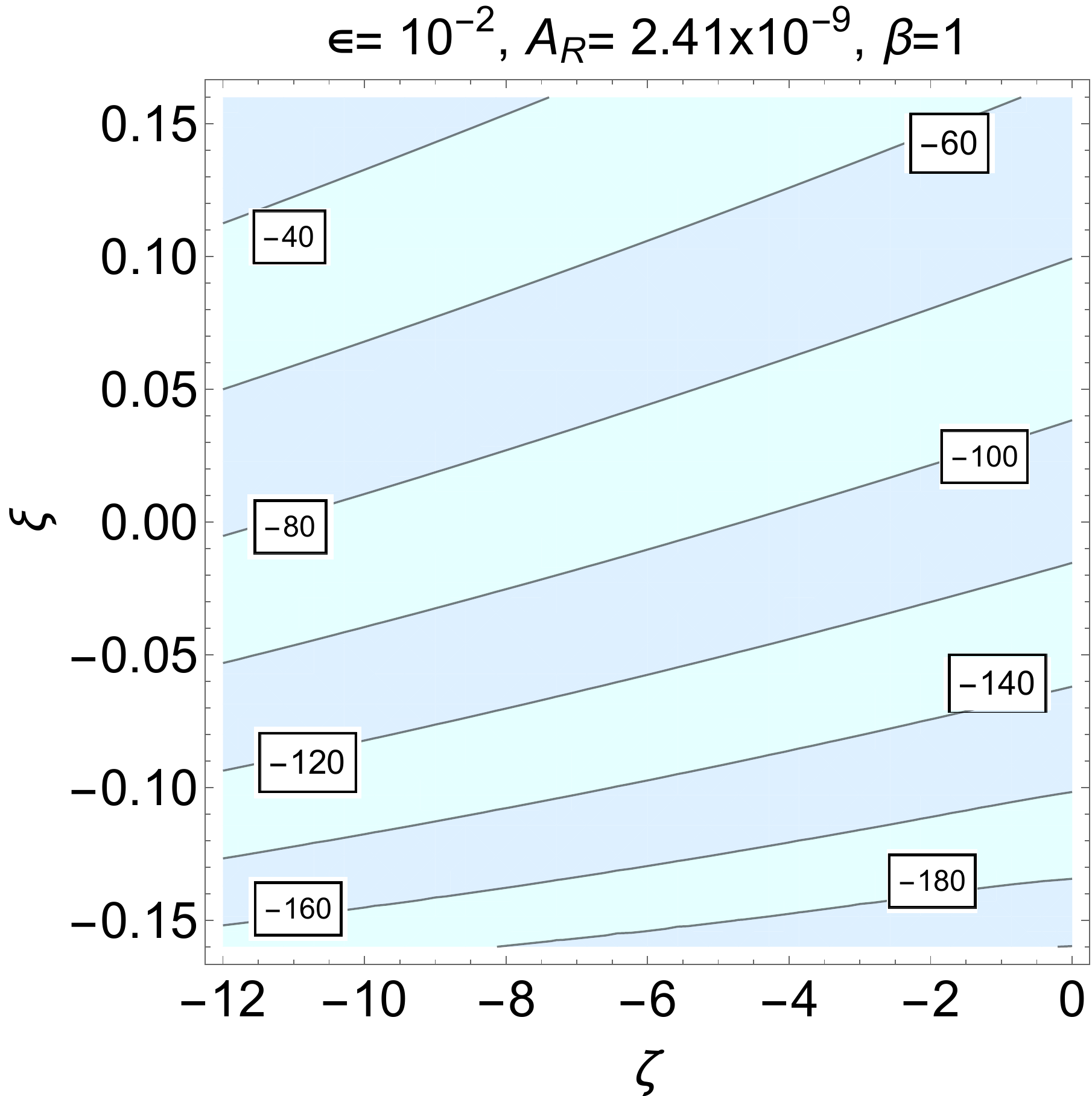}
\includegraphics[height=7cm]{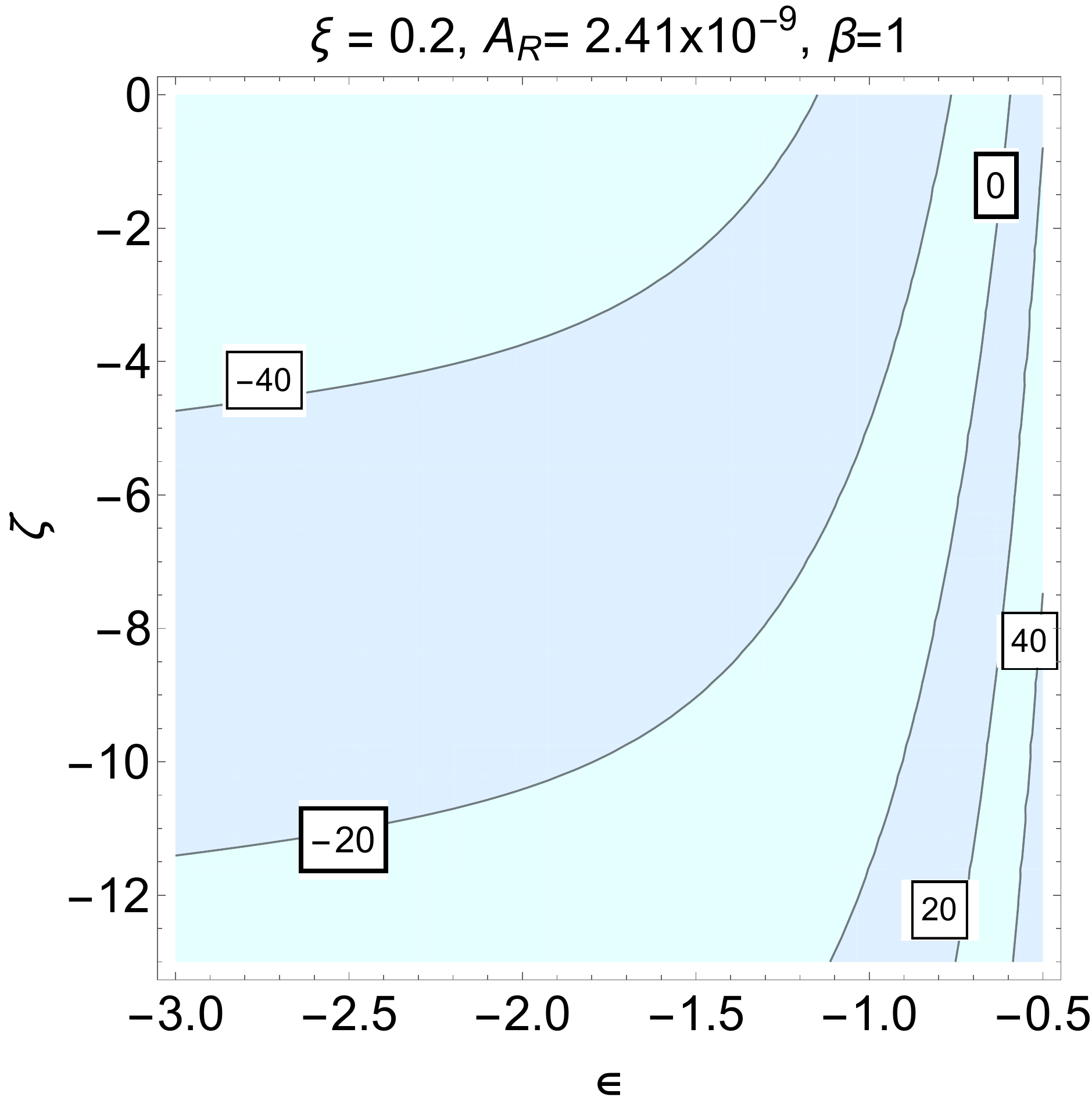}
\caption[a]{The parameter space is projected on four complementary planes at the typical scale of the protogalactic collapse. The 
labels appearing on the curves denote the common logarithms of $\Omega_{B}$. Common logarithms are used on all the axes except for $\xi$ which 
is illustrated on a linear scale; $k_{L} = 1/\mathrm{Mpc}$. }
\label{Figure1}      
\end{figure}

Even if accurate analyses of the numerical value of the ratio $(\sigma_{c}/T)$ have been discussed \cite{max4} the most relevant point is that the conductivity 
increases linearly with the temperature. Thus the ratio $\sigma_{c}/H_{e}$ determines $k_{\sigma}$ and can then be explicitly estimated from:
\begin{equation}
\frac{T}{H_{e}} = \biggl(\frac{135}{4\pi^4 N_{eff}}\biggr)^{1/4} (\epsilon\,{\mathcal A}_{{\mathcal R}})^{-1/4},
\label{MM2}
\end{equation}
where $N_{eff}$ denotes the effective number of spin degrees of freedom (in concrete estimates we shall always assume  $N_{eff} = 106.75$ , as it happens in the standard model). In the sudden reheating approximation the plasma right after inflation is in thermal equilibrium  at $T$, up to small  perturbations of the distribution functions leading to charge and current fluctuations. As the Universe decelerates after inflation the ratio of the 
 conductivity to the Hubble rate increases. 
 \begin{figure}[!ht]
\centering
\includegraphics[height=7cm]{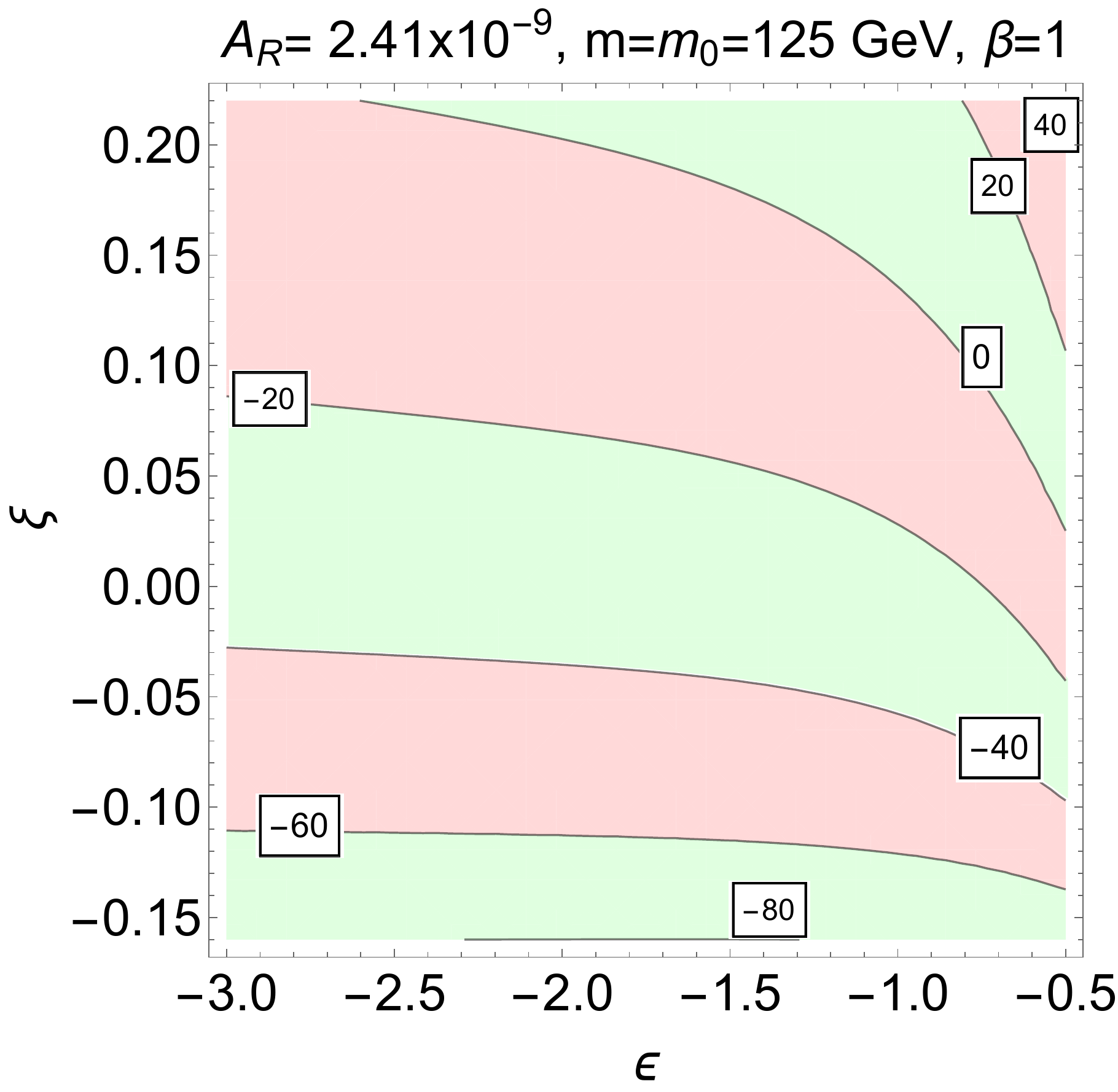}
\includegraphics[height=7cm]{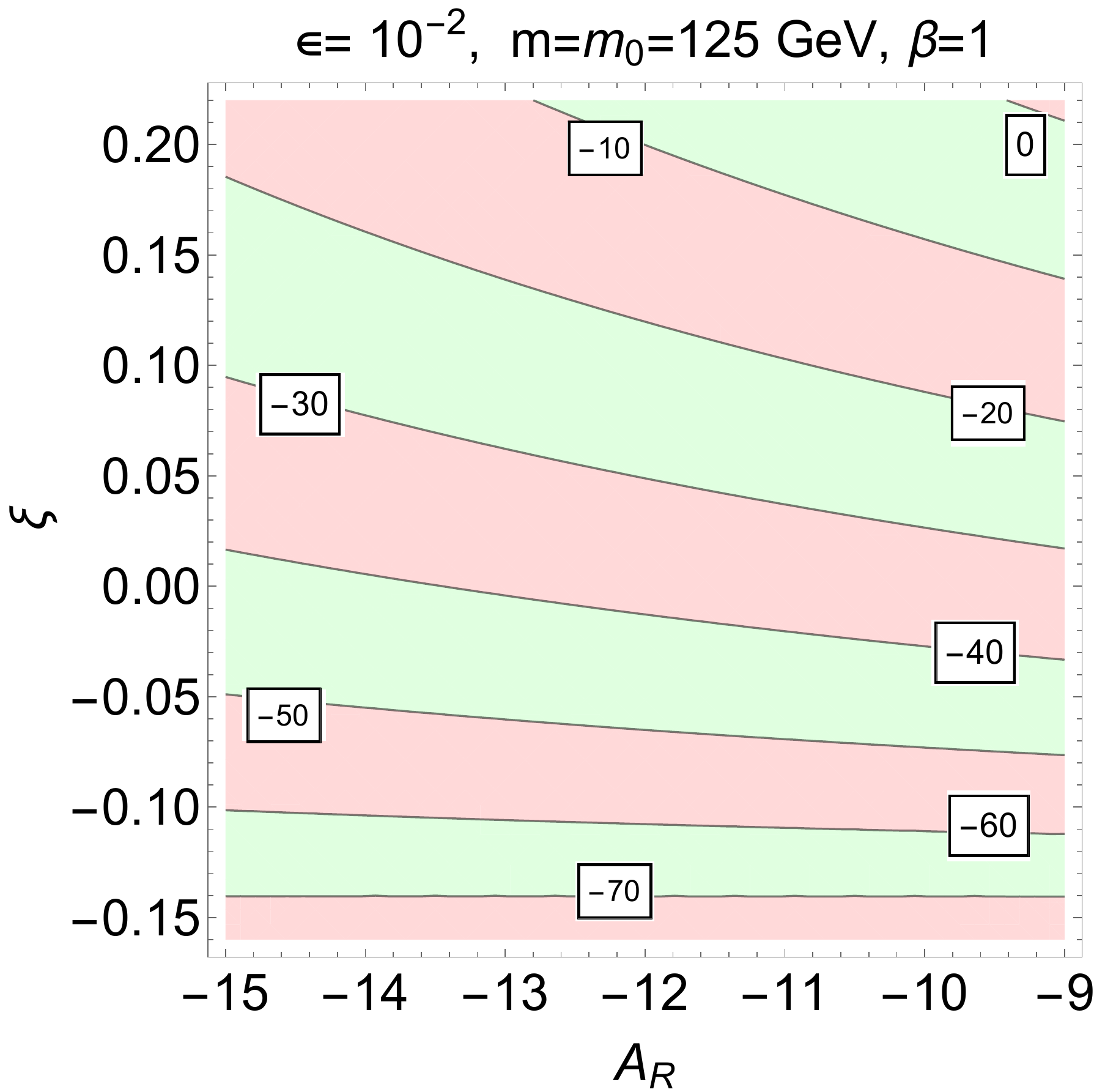}
\includegraphics[height=7cm]{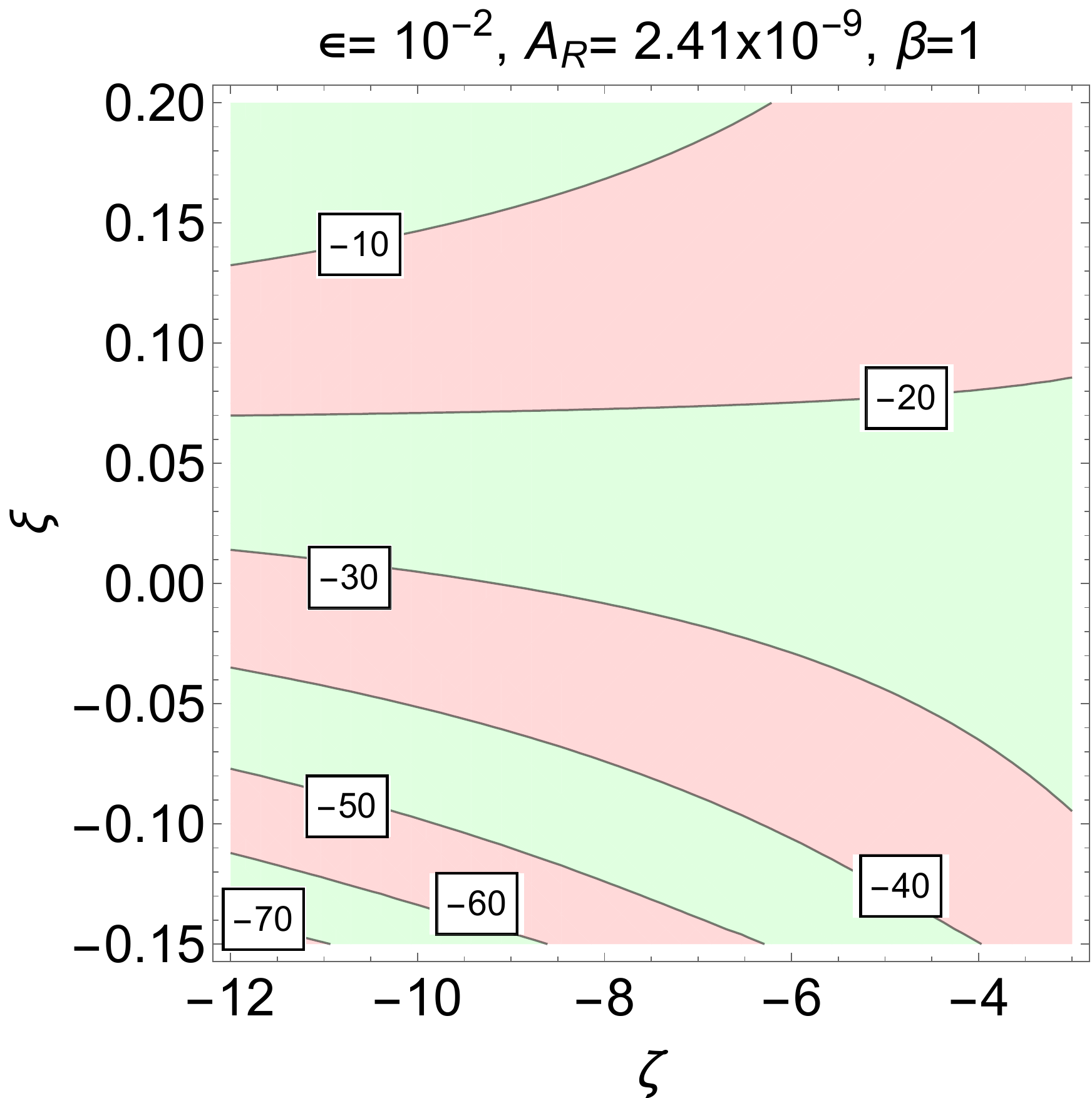}
\includegraphics[height=7cm]{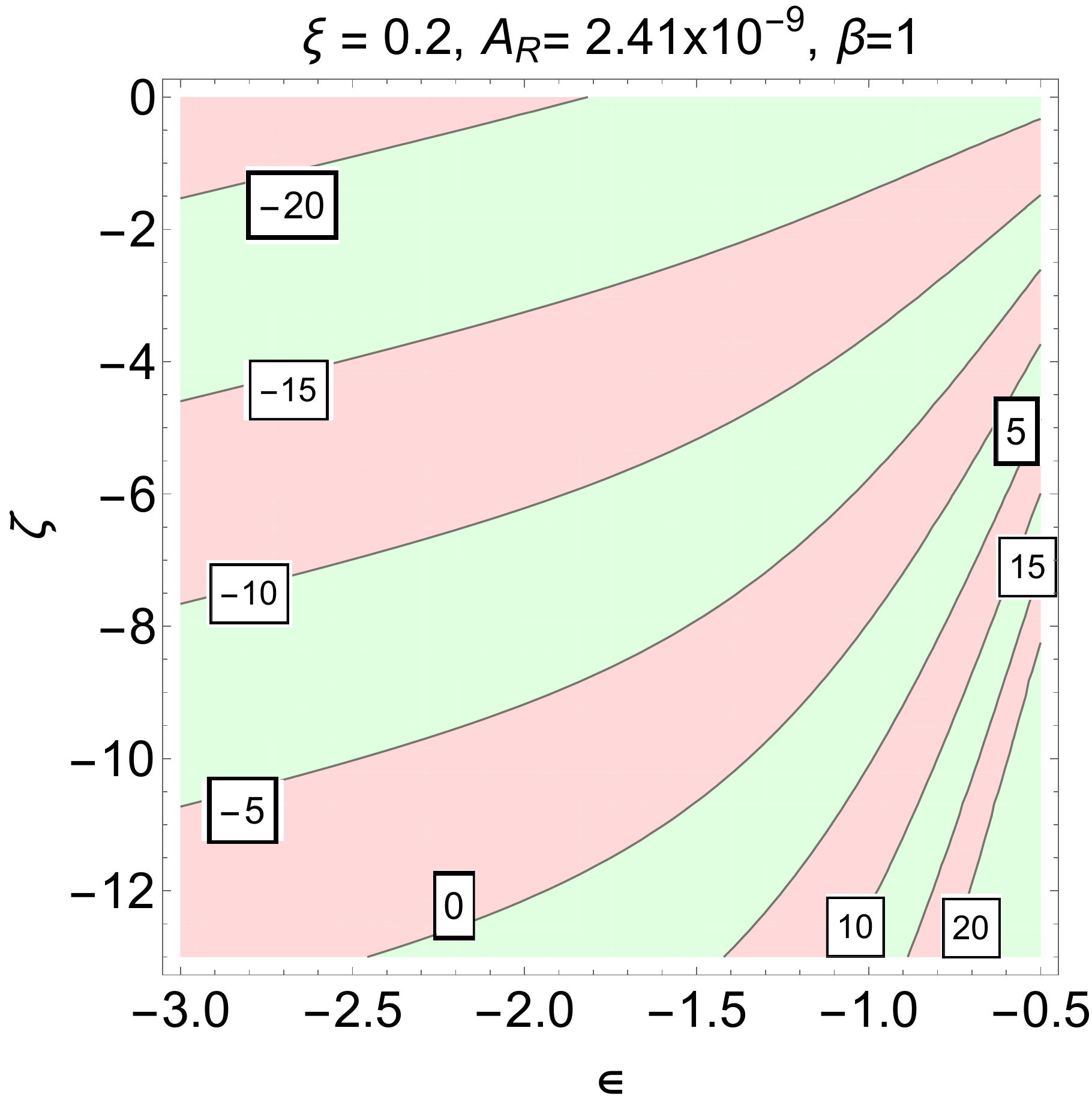}
\caption[a]{We illustrate the parameter space when $k_{L}(a_{e} H_{e}) = \zeta$
which coincides with the largest wavenumber of the problem. The same notations adopted in Fig. \ref{Figure1} have been 
consistently employed; as in Fig. \ref{Figure1}, $k_{L} = 1/\mathrm{Mpc}$.}
\label{Figure2}      
\end{figure}
Before proceeding further it is interesting to note that the {\em spectral energy density} (i.e. the energy density per logarithmic interval of wavenumber) coincides 
 with the magnetic power spectrum. The Fourier transform of the magnetic field intensity obeys 
\begin{eqnarray}
B_{i}(\vec{x},\tau) &=& \frac{1}{(2 \pi)^{3/2}} \int d^{3} k B_{i}(\vec{k},\tau) \, e^{- i \vec{k}\cdot \vec{x}},
\nonumber\\
\langle B_{i}(\vec{k},\tau) \, B_{j}(\vec{p},\tau) \rangle &=& \frac{2\pi^2}{k^3} P_{ij}(\hat{k}) P_{B}(k,\tau) \delta^{(3)}(\vec{k}+\vec{p})
\label{PS}
\end{eqnarray}
 where $P_{B}(k,\tau) $ is the magnetic power spectrum and $P_{ij}(\hat{k}) = (\delta_{ij} -\hat{k}_{i} \hat{k}_{j})$ the traceless projector. 
 Note that $P_{B}(k,\tau)$ has dimensions of an energy density. Let us now formally compute the averaged magnetic energy density; we will have from Eq. (\ref{PS}) 
 \begin{equation}
 \rho_{B} = \frac{1}{2 a^4} \langle B^2\rangle = \frac{1}{a^4}\int \frac{d k}{k} P_{B}(k,\tau).
 \label{PS2}
 \end{equation}
 where we used Eq. (\ref{PS}). Equation (\ref{PS2}) defines the spectral energy density 
 which we ought to express in  units of the energy density of radiation 
 \begin{equation} 
 \Omega_{B} = \frac{1}{\rho_{r}} \frac{d \rho_{B}}{d \ln{k}} \equiv \frac{P_{B}(k,\tau)}{a^4 \rho_{r}},
 \label{PS3}
 \end{equation}
 implying that the slope of the spectral energy density coincides with the slope of the magnetic power spectrum defined according to Eq. (\ref{PS2}). Since the numerator and the denominator scale in the same way in Eq. (\ref{PS3}), $\Omega_{B}$ will be approximately constant during the radiation epoch. During the matter epoch 
 $\Omega_{B}$ will still be constant but will not coincide anymore with the magnetic energy density in critical units. 
  The results of Eqs. (\ref{nu}) and  (\ref{CC4})--(\ref{ChR}), once inserted into Eq. (\ref{BB}), determine $\Omega_{B}$ through Eq. (\ref{PS3}): 
\begin{eqnarray}
&&\Omega_{B}(k_{L}, \xi, \zeta, \epsilon, {\mathcal A}_{{\mathcal R}},\beta) = (\epsilon {\mathcal A}_{{\mathcal R}})^{7/2}\,{\mathcal S}(\alpha,\theta_{W}) \, {\mathcal T}(\alpha,\beta, \epsilon, \zeta, {\mathcal A}_{{\mathcal R}}) \, \, \biggl(\frac{k_{L}}{a_{e} H_{e}} \biggr)^{ 10 - 4 \alpha} \biggl(\frac{m}{M_{P}}\biggr)^{-3},
\nonumber\\
&& {\mathcal T}(\alpha,\beta, \epsilon, \zeta,{\mathcal A}_{{\mathcal R}}) =\frac{9 N_{eff}^{3/2-\alpha} }{10\,\beta^{3}\, \overline{\alpha}}\biggl[ (1 - 2\alpha + 2 \beta)^2\zeta \Gamma^2(3/4) + 4 ( 1 - 2\alpha)^2 \beta \Gamma^2(5/4) \biggr]^2,
\nonumber\\
&& {\mathcal S}(\alpha,\theta_{W}) = 2^{2 \alpha - 1} \,5^{\alpha-3/2}\,3^{2 \alpha -4} \,\pi^{7/2 - 2 \alpha} \,(6 - 4 \sqrt{2}) \,\Gamma^4(\alpha)\, \cos^2{\theta}_{W},
\label{OMB}
\end{eqnarray}
which is valid for $k_{L}/(a_{e} H_{e}) \leq m/H_{e} \equiv \zeta$. According to Eq. (\ref{MM2}) $\zeta \ll k_{\sigma}/(a_{e} H_{e})$ and $k_{\sigma}/(a_{e} H_{e})
= {\mathcal O}(1)$.
If $k_{L}$ is evaluated at the time of the collapse of the protogalaxy we have instead that:
\begin{equation}
\frac{k_{L}}{a_{e} H_{e}} = \frac{k_{L}}{H_{0} } e^{ - N_{{max}}},\qquad e^{N_{{max}}} = ( 2 \, \pi  \, \epsilon \, {\mathcal A}_{{\mathcal R}}\, \Omega_{\mathrm{R}0})^{1/4} \,\, \biggl(\frac{M_{\mathrm{P}}}{ H_{0} }\biggr)^{1/2} \biggl(\frac{H_{r}}{H_{e}}\biggr)^{w-1/2},
\label{xx}
\end{equation}
where $H_{0} =2.334\times 10^{-4} \,(h_{0}/0.7) \mathrm{Mpc}^{-1}$  is the present value of the Hubble rate and 
$\Omega_{\mathrm{R}0}$ is the present critical fraction of radiation (in the concordance model 
$h_{0}^2 \Omega_{\mathrm{R}0} = 4.15 \times 10^{-5}$).

According to Eq. (\ref{OMB}) the scale-invariant limit of the spectral energy density is obtained for $\alpha \simeq 5/2$.  By definition 
this will also correspond to the scale-invariant limit of the power spectrum. This specific value of $\alpha$ correspond to a specific value of $\xi$. 
More specifically from Eq. (\ref{BB3a}) we will have that $\xi$ is determined, approximately, by requiring 
\begin{equation}
\frac{3}{2} \sqrt{1 + \frac{16}{3} \xi - \frac{\zeta^2}{9} } \simeq \frac{5}{2}. 
\end{equation}
But since $\zeta \ll 1$ we will have that the scale-invraint limit $\alpha\to 5/2$ is achieved when $\xi \simeq 1/3$. As we shall see  this value 
is too large and it is incompatible with the bounds discussed in Fig. \ref{Figure1} (see in particular the $(\epsilon, \, \xi)$ plane).

In Eq. (\ref{xx})  $N_{{max}}$ denotes the maximal number of efolds which are today accessible to our observations  \cite{mgg}:
in practice $N_{{max}}$ is determined by fitting the redshifted inflationary event horizon inside the present 
Hubble radius $H_{0}^{-1}$; the term containing $w$ accounts for the possibility of a delayed reheating 
ending at the scale $H_{r}$ eventually much smaller than the inflationary curvature scale $H_{e}$. 
For illustration we shall choose the sudden reheating approximation by setting $w =1 /2$ but different possibilities have been considered in the 
literature\footnote{A long postinflationary phase dominated by a stiff equation of state has been examined in the context of magnetogenesis 
(see first paper of Ref. \cite{mgg} and references therein). A delayed reheating has the effect of increasing $N_{max}$. The largest 
value of $N_{max}$ in the case of a stiff postinflationary phase can be estimated as $N_{max} = 78.3 + (1/3) \ln{\epsilon}$.}.
In the sudden reheating approximation we have  $N_{{max}} \simeq 63.25 + 0.25 \ln{\epsilon}$ which is 
numerically close to the minimal number of efolds needed to solve the kinematic 
problems of the standard cosmological model.  Both in Figs. \ref{Figure1} and \ref{Figure2} we have $\beta=1$. Values $\beta \neq 1$ parametrize the deviations
from the slow-roll dynamics. This situation might be interesting in its own right but it is not central to the present discussion\footnote{The case when $\epsilon$  is not constant 
will have to reproduce the same quantitive features of the constant epsilon case: since 
the external background is fixed one can always compute the value of epsilon 50 or 60 efolds 
before the end of inflation, compute the magnetic field and discuss the various bounds.}.

According to Eq. (\ref{BB3}) $\alpha = \alpha(\epsilon, \xi, \zeta)$ so that $\alpha$ is not an independent parameter. The parameter space is therefore 
four-dimensional and it involves $\xi$, $\epsilon$, $\zeta$, and ${\mathcal A}_{{\mathcal R}}$. The variation 
of ${\mathcal A}_{{\mathcal R}}$ is used to examine the regions where $H_{e}$ is artificially low in Planck units (recall, in this 
respect, Eqs. (\ref{sc1}))--(\ref{sc2}).  In principle there could be a further parameter, namely $\beta$. Values $\beta \neq 1$ parametrize the deviations
from the slow-roll dynamics. This situation might be interesting in its own right but it is not central to the present discussion. We shall therefore 
focus on the situation $\beta=1$ where, incidentally, the term $(1- 2 \alpha + 2\beta)$ is suppressed in Eq. (\ref{OMB}) 
for $\xi \to 0$. In Fig. \ref{Figure1}  the magnetic energy density in critical units is illustrated at the typical pivot scale of the protogalactic collapse (i.e. 
$k_{L} =  \mathrm{Mpc}^{-1}$). 
Except for $\xi$, all the other quantities on the horizontal and vertical axes (i.e. ${\mathcal A}_{R}$, 
$\epsilon$ and $\zeta$) are reported in terms of the corresponding common logarithms. In Fig. \ref{Figure2} the parameter space is instead 
charted for $k_{L}/(a_{e} H_{e}) = \zeta$. It is useful to check, from Figs. \ref{Figure1} and \ref{Figure2},  that in the minimally coupled case (i.e. $\xi\to 0$) the constraints are all automatically satisfied since $\Omega_{B}$  ranges from ${\mathcal O}(10^{-80})$ to ${\mathcal O}(10^{-35})$. The minimally coupled limit of Eq. (\ref{minimal}) corresponds to the intercept $\xi=0$ in the various plots of Figs. \ref{Figure1} 
and \ref{Figure2}. In the limit $\xi\to 0$, up to slow-roll corrections,
 $\alpha \to 3/2$ and Eq. (\ref{OMB}) implies:
\begin{equation}
\Omega_{B} \simeq 1.61\times 10^{-72} \biggl(\frac{k_{L}}{\mathrm{Mpc}^{-1}}\biggr)^4 \cos^2{\theta_{W}} \biggl(\frac{m}{\mathrm{GeV}}\biggr)^{-3},
\label{minimal}
\end{equation}
that coincides with the previous results of Ref. \cite{mg2} where $B^2/T^4 \simeq 10^{-80}$ for a typical mass $m= 100 \,\mathrm{GeV}$.  At the maximal scale $k_{L}/(a_{e} H_{e}) = \zeta$ (and in the limit $\xi \to 0$) the magnetic energy density in critical units is instead given by $6.54\cos^2{\theta_{W}} (m/M_{P})({\mathcal A}_{{\mathcal R}} \epsilon)^{3/2}$ which is never overcritical. As previously established in the case $\xi \to 0$ there are practically no constraints.
\begin{figure}[!ht]
\centering
\includegraphics[height=7cm]{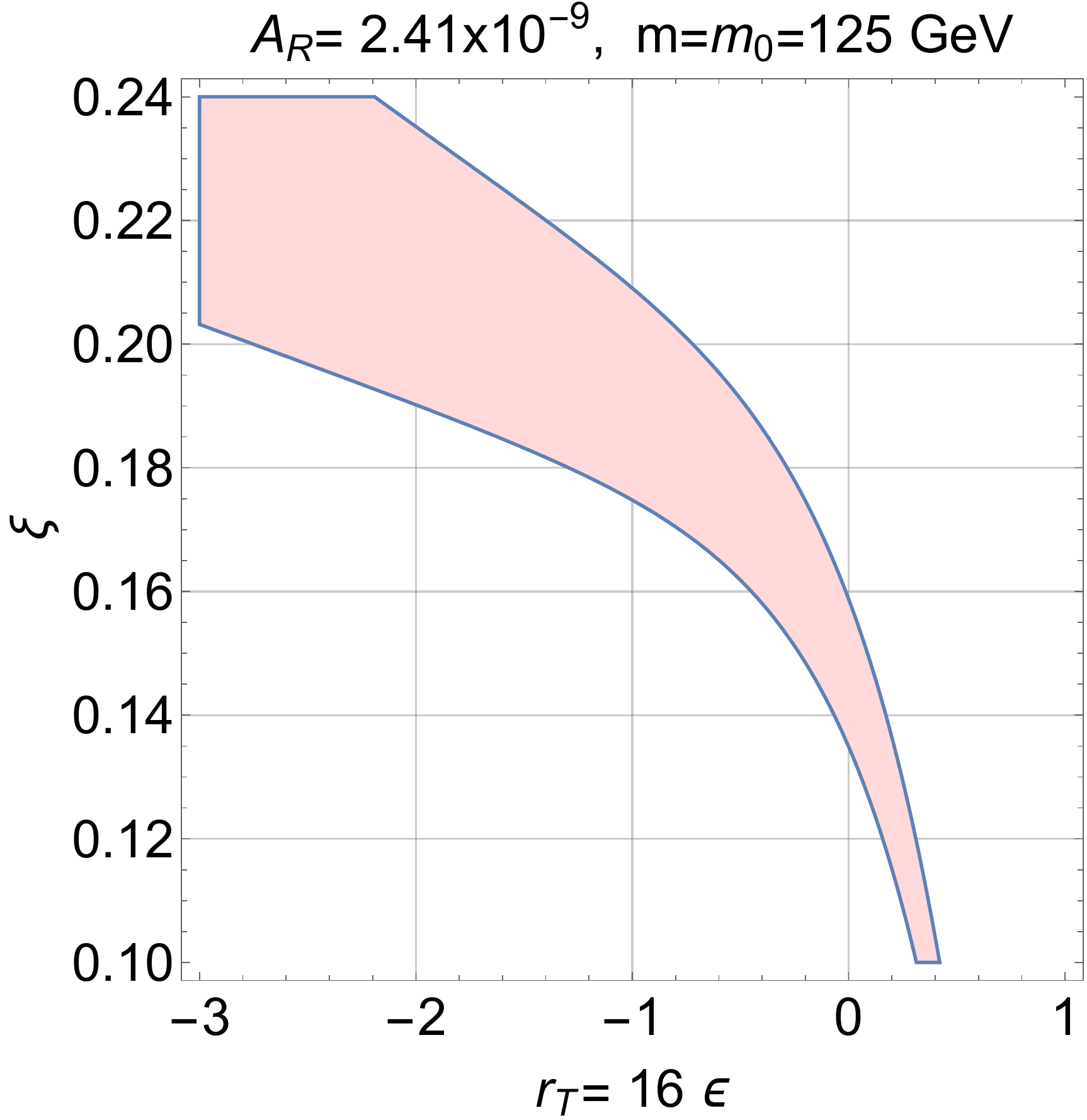}
\includegraphics[height=7cm]{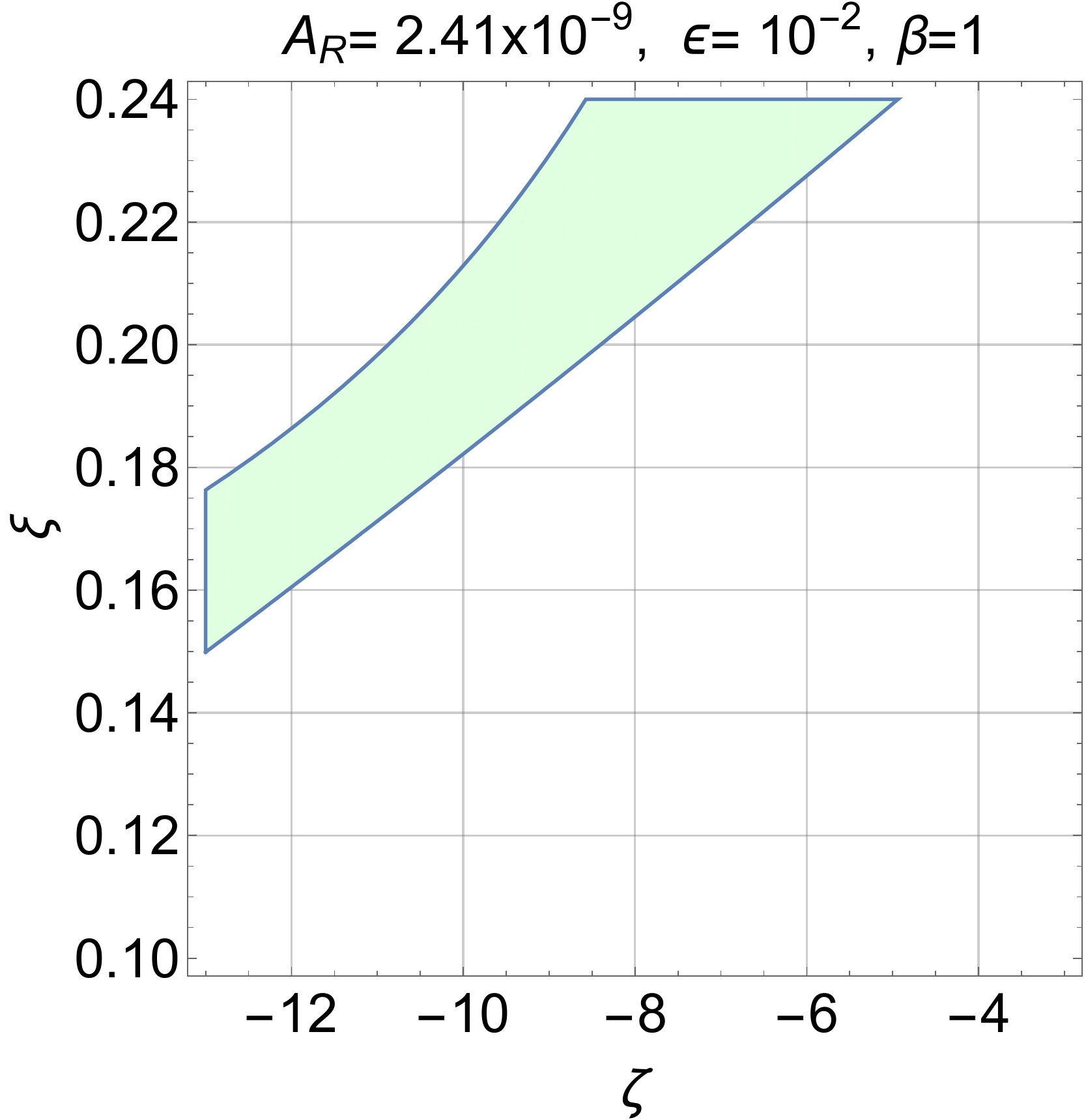}
\caption[a]{We illustrate the magnetogenesis constraints in the planes $(\xi,\,\epsilon)$ and $(\xi,\, \zeta)$. The filled areas denote the allowed regions.}
\label{Figure3}      
\end{figure}
From the $(\xi, \epsilon)$ plane (plot at the top left in Fig. \ref{Figure1}) the closure bound is violated for $\xi > {\mathcal O}(0.25)$ and 
$\epsilon < {\mathcal O}(10^{-3})$. The $(\xi, {\mathcal A}_{{\mathcal R}})$ plane illustrates, as already mentioned, the situation where the inflationary scale 
is artificially lowered as in \cite{low}. The consistency relations do not necessarily hold in this case but the values of the magnetic fields are even smaller and there are 
no constraints on $\xi$. When charting the $(\xi, {\mathcal A}_{{\mathcal R}})$ plane 
we set the effective mass to its minimal value which is the most constraining: larger values of the mass will suppress the magnetic energy density 
even further. A similar conclusion emerges from  the analysis of the $(\xi,\zeta)$ plane.

If the magnetic energy density is approximately larger than $10^{-22}\,\, \mathrm{nG}^2$ at the scale of the protogalactic collapse, the observed magnetic fields could be  
subsequently amplified by the combined action of the gravitational collapse itself and by the galactic rotation 
which transforms the kinetic energy of the plasma into magnetic energy (this is the meaning of the so-called galactic dynamo \cite{parker} which has a long history \cite{mg3}). The most optimistic estimates for the required initial conditions are derived by assuming that every rotation of the galaxy would increase the magnetic field of one efold. The number of galactic rotations since the collapse of the protogalaxy can be between $30$ and $35$,  leading approximately to  a purported growth of $13$ orders of magnitude (see \cite{mg3} and discussions therein).  If the dynamo action is totally absent the required field should be ${\mathcal O}(10^{-2}) \mathrm{nG}$ which means $\Omega_{B} = {\mathcal O}(10^{-10})$. In this case during the collapse of the protogalaxy the magnetic field will increase of about $5$ orders of magnitude.
The shaded regions in Fig. \ref{Figure3} illustrate the interplay between the closure bounds and the magnetogenesis 
requirements. The seed field at the onset of the gravitational collapse must be, at least, as large as $10^{-15}$ nG or, more realistically, larger than $10^{-11}$ nG.  
The simplest way to implement the magnetogenesis requirements is to plot the magnetic energy density
for all the regions where the backreaction constraints are satisfied. The areas of the parameter 
space where $  \log{\Omega_{B}} > - 25$ (or, even more optimistically,  $  \log{\Omega_{B}} > - 35$) will therefore offer viable models of magnetogenesis. 

In Fig. \ref{Figure3} we illustrate the region where the closure bounds are satisfied and $\Omega_{B} > 10^{-25}$.
In these regions the magnetogenesis requirements are met for $0.2\leq \xi < 0.24$  and $\epsilon < 0.01$. 
The allowed regions of Fig. \ref{Figure3} shrink almost to a point if we enforce the closure bound and demand, 
at the scale of the protogalactic collapse, $\log{\Omega_{B}} > -14$.
If we impose an even more demanding constraint (i.e. $\log{\Omega_{B}} > -11$) 
the allowed regions disappear 
completely. If $\Omega_{B} = {\mathcal O}(10^{-11})$ the magnetic field of the galaxy could be obtained 
almost completely by compressional amplification \cite{mg3}. 
Compressional amplification alone is therefore insufficient and to obtain the observed fields 
some sort of dynamo action is mandatory, at least in the present framework. 
A more detailed account of these themes is beyond the scopes of this paper but can 
be found in more comprehensive reviews describing the features of our magnetized universe \cite{mg3} 
(see, in particular, first and second reviews).  
Note finally that direct constraints can be obtained from CMB data 
(see, for instance, \cite{ref2} for some recent analyses obtained from the temperature and polarization anisotropies). Indeed, as noticed over ten years ago, 
the temperature and the polarization anisotropies of the CMB may be magnetized since the initial conditions of the scalar modes are affected by the presence 
of stochastic magnetic fields \cite{magnetized1}. This observation offers the unique opportunity of direct limits on the large-scale magnetism 
prior to matter-radiation equality since the large-scale magnetic fields affect directly the initial conditions of the Einstein-Boltzmann hierarchy. 
The current Planck explorer data have been used to set bounds on large-scale magnetic fields coming from inflation as previoulsy done with the WMAP 
3-yr and 9-yr releases (see, respectively, third and fourth papers of Ref.  \cite{magnetized1}).  
The allowed regions of Fig. \ref{Figure3} are abundantly compatible with the CMB constraints. 

\renewcommand{\theequation}{5.\arabic{equation}}
\setcounter{equation}{0}
\section{Concluding remarks} 
\label{sec5}
We analyzed the parametric amplification of massive Higgs excitations when the coupling to gravity 
is non-minimal. In the framework of the Abelian-Higgs model this processes may lead to overcritical large-scale 
gauge fields.  As a consequence the non-minimal coupling cannot be too large (i.e. typically $\xi \leq 0.25$) when 
 $\epsilon < {\mathcal O}(10^{-2})$. When the coupling is either minimal (i.e. $\xi \to 0$) or conformal (i.e. $\xi \to -1/6$) 
there are practically no constraints. A global scan of the parameter space suggests  that the large-scale magnetic fields 
produced when $\xi = {\mathcal O}(0.2)$ turn out to be partially relevant for the problem of 
primordial magnetogenesis. 

\section*{Acknowledgments} 
It is a pleasure to thank A. Gentil-Beccot and S. Rohr for their kind help.


\newpage 

\begin{thebibliography}{99}

\bibitem{WMAP1} D.~N.~Spergel {\it et al.},   Astrophys.\ J.\ Suppl.\  {\bf 148}, 175 (2003); D.~N.~Spergel {\it et al.},
{\em ibid.} \ {\bf 170}, 377 (2007);  L.~Page {\it et al.} {\em ibid.}  {\bf 170}, 335 (2007).

\bibitem{WMAP2} B.~Gold {\it et al.},  Astrophys.\ J.\ Suppl.\ {\bf 192}, 15 (2011); 
D.~Larson,  {\it et al.},  {\em ibid.}  {\bf 192}, 16 (2011); C.~L.~Bennett {\it et al.}, {\em ibid.}\ {\bf 192}, 17 
(2011); G.~Hinshaw {\it et al.},  {\em ibid.} {\bf 208} 19 (2013); C.~L.~Bennett {\it et al.},   {\em ibid.} {\bf 208} 20 (2013); 
 P.~A.~R.~Ade {\it et al.}  [Planck Collaboration],   Astron.\ Astrophys.\  {\bf 571}, A22 (2014); Astron.\ Astrophys.\  {\bf 571}, A16 (2014).

\bibitem{ACBARQUAD} C.~L.~Reichardt, P.~A.~R.~Ade, J.~J.~Bock {\it et al.}, Astrophys.\ J.\  {\bf 694}, 1200-1219 (2009);
M.~Zemcov {\it et al.}  [QUaD collaboration], Astrophys.\ J.\  {\bf 710}, 1541 (2010).

\bibitem{hh} K.~Enqvist, H.~Kurki-Suonio and J.~Valiviita, Phys.\ Rev.\  D {\bf 62}, 103003 (2000);  
J.~Valiviita and V.~Muhonen, Phys.\ Rev.\ Lett.\  {\bf 91}, 131302 (2003); 
H.~Kurki-Suonio, V.~Muhonen and J.~Valiviita, Phys.\ Rev.\  D {\bf 71}, 063005 (2005); 
M.~Giovannini,  Class.\ Quant.\ Grav.\  {\bf 23}, 4991 (2006).

\bibitem{LSS} W.~J.~Percival, B.~A.~Reid, D.~J.~Eisenstein {\it et al.},  Mon.\ Not.\ Roy.\ Astron.\ Soc.\  {\bf 401}, 2148-2168 (2010); 
B.~A.~Reid, W.~J.~Percival, D.~J.~Eisenstein {\it et al.}, Mon.\ Not.\ Roy.\ Astron.\ Soc.\  {\bf 404}, 60 (2010).

\bibitem{SNN} R.~Kessler, A.~Becker, D.~Cinabro {\it et al.},  Astrophys.\ J.\ Suppl.\  {\bf 185}, 32 (2009);  M.~Hicken, W.~M.~Wood-Vasey, S.~Blondin {\it et al.}, Astrophys.\ J.\  {\bf 700}, 1097 (2009).

\bibitem{HG} G. Aad et al. [ATLAS Collaboration], Phys. Lett. B {\bf 716}, 1 (2012);  S. Chatrchyan et al. [CMS Collaboration], Phys. Lett. B {\bf 716}, 3 (2012).

\bibitem{k1} K.~Enqvist, T.~Meriniemi and S.~Nurmi, JCAP {\bf 1310}, 057 (2013); {\em ibid}. {\bf 1407}, 025 (2014).

\bibitem{low} M.~Giovannini,  Phys.\ Rev.\ D {\bf 67}, 123512 (2003);  D.~H.~Lyth,
  Phys.\ Lett.\ B {\bf 579}, 239 (2004); Phys.\ Rev.\ D {\bf 69}, 083509 (2004).

\bibitem{tur} M.S. Turner, Phys. Rev. D {\bf 28}, 1243  (1983).

\bibitem{ch} K.~Y.~Choi and Q.~G.~Huang,  Phys.\ Rev.\ D {\bf 87}, no. 4, 043501 (2013).

\bibitem{low2} M.~Giovannini,   Phys.\ Rev.\ D {\bf 58}, 083504 (1998); Phys.\ Rev.\ D {\bf 60}, 123511 (1999);
 Class.\ Quant.\ Grav.\  {\bf 26}, 045004 (2009).

\bibitem{mod}  D.~H.~Lyth, JCAP {\bf 0511}, 006 (2005);  Phys.\ Rev.\ Lett.\  {\bf 97}, 121301 (2006).
  
\bibitem{rec} M.~Herranen, T.~Markkanen, S.~Nurmi and A.~Rajantie,  Phys.\ Rev.\ Lett.\  {\bf 115}, 241301 (2015).

\bibitem{SO} T. Muta and S.D. Odintsov,  Mod. Phys. Lett. A {\bf 6} 3641 (1991); 
S. Mukaigawa, T. Muta,  and S.D. Odintsov,  Int. J. Mod. Phys.  A {\bf 13}  2739 (1998).

\bibitem{mg0} M. Giovannini,   Phys.\ Lett.\ B {\bf 659}, 661 (2008);  Phys.\ Rev.\ D {\bf 85}, 101301 (2012); Phys.\ Rev.\ D {\bf 86}, 103009 (2012).

\bibitem{mg2} M.~Giovannini and M.~E.~Shaposhnikov, Phys.\ Rev.\ D {\bf 62}, 103512 (2000).

\bibitem{mg3} K. Enqvist, Int. J. Mod. Phys. D {\bf 07}, 331 (1998); M. Giovannini, 
Int. J. Mod. Phys. D {\bf 13}, 391 (2004); J.D. Barrow, R. Maartens, and C. G. Tsagas, Phys. Rep. {\bf 449}, 131 (2007).

\bibitem{chrom}  A. Linde, Phys. Lett. B {\bf 96}, 289 (1980); 
D.J. Gross, R.D. Pisarski and L.G. Yaffe, Rev. Mod. Phys. {\bf 53}, 43 (1981).

\bibitem{parker} E. N. Parker, {\it Cosmical Magnetic Fields} (Clarendon Press, Oxford, 1979).

\bibitem{staro} Ya. B. Zel'dovich and A. A. Starobinsky, Zh. Eksp. Teor. Fiz.
{\bf 61}, 2161 (1971) [Sov. Phys. JETP {\bf 34}, 1159 (1972)]; Pis'ma Zh.
Eksp. Teor. Fiz. {\bf 26}, 373 (1977) [JETP Lett. 26, 252 (1977)].

\bibitem{birrel} N. D. Birrell and P. C. W. Davies, {\it Quantum Fields in Curved Space} (Cambridge University
Press, Cambridge, 1982).

\bibitem{ford1} L. H. Ford, Phys. Rev. D {\bf 35}, 2955 (1987); E.~D.~Schiappacasse and L.~H.~Ford,  arXiv:1602.08416 [gr-qc].

\bibitem{ford2} L.~H.~Ford,  Phys.\ Rev.\ D {\bf 31}, 704 (1985).

\bibitem{assi} M.~Giovannini,  Class.\ Quant.\ Grav.\  {\bf 20}, 5455 (2003).

\bibitem{mollow}  B. L. Mollow and R. J. Glauber, Phys. Rev. {\bf 160}, 1076 (1967); {\it ibid}. {\bf 160}, 1097 (1967); 
D.~Stoler,  Phys.\ Rev.\  D {\bf 1}, 3217 (1970);  Phys.\ Rev.\  D {\bf 4}, 1925 (1971).

\bibitem{mb}  A.L.~Fetter and J.D.~Walecka, {\it Quantum Theory of Many-Particle Systems} (McGraw-Hill, New York, 1971);
 A. I. Solomon J.\ Math.\ Phys.\  {\bf 12}, 390 (1971).

\bibitem{abr}  A. Erdelyi, W. Magnus, F. Oberhettinger, and F. R. Tricomi {\em Higher Trascendental Functions} (Mc Graw-Hill, New York, 1953); 
M. Abramowitz and I. A. Stegun, {\em Handbook of Mathematical Functions} (Dover, New York, 1972).

\bibitem{pl1}  T. J. M Boyd and J. J. Sanderson, {\it The physics of plasmas}, (Cambridge University Press, Cambridge, England, 2003). 

\bibitem{pl2} J. Bernstein, {\it Kinetic Theory in the Expanding Universe}, (Cambridge University Press, Cambridge, England, 1988).
 
\bibitem{max4} J. Ahonen, Phys. Rev. D {\bf 59}, 023004 (1998); J. Ahonen and K. Enqvist, Phys. Lett. B {\bf 382}, 40 (1996); 
H. Heiselberg, Phys. Rev. D {\bf 49}, 4739 (1994).

\bibitem{DT1} B. Ratra,  Astrophys.\, J.\, Lett.  {\bf 391}, L1 (1992); M.~Gasperini, M.~Giovannini, and G.~Veneziano, Phys. Rev. Lett. {\bf 75}, 3796 (1995); 
M.~Giovannini,  Phys.\ Rev.\ D {\bf 56}, 3198 (1997); M.~Giovannini,  Phys.\ Rev.\  D {\bf 64}, 061301 (2001).
 
\bibitem{DT2} K.~Bamba and M.~Sasaki,  JCAP {\bf 02}, 030 (2007); K. Bamba JCAP {\bf 10}, 015 (2007); M. Giovannini, Phys.\ Lett.\  B {\bf 659}, 661 (2008).

\bibitem{DT3} K. Bamba, Phys. Rev. D {\bf 75} 083516 (2007); J.~Martin and J.~'i.~Yokoyama,ÊÊJCAP {\bf 0801}, 025 (2008); M. Giovannini, Lect.\ Notes Phys.\  {\bf 737}, 863 (2008);
 S.~Kanno, J.~Soda and M.~-a.~Watanabe, ÊÊJCAP {\bf 0912}, 009 (2009); L. Campanelli, Phys.\ Rev.\ D {\bf 93}, no. 6, 063501 (2016); L.~Campanelli and A.~Marrone,
  Phys.\ Rev.\ D {\bf 94}, no. 10, 103510 (2016).
  
\bibitem{VdW}   M.~Giovannini,  Phys.\ Rev.\ D {\bf 88}, no. 8, 083533 (2013);  Phys.\ Rev.\ D {\bf 92}, no. 4, 043521 (2015); Phys.\ Rev.\ D {\bf 92}, no. 12, 121301 (2015);
G.~Tasinato,  JCAP {\bf 1503}, 040 (2015); R.~Z.~Ferreira and J.~Ganc,  JCAP {\bf 1504}, no. 04, 029 (2015);  M.~Giovannini,
   Phys.\ Rev.\ D {\bf 93}, no. 4, 043543 (2016).
    
\bibitem{mgg} M.~Giovannini,  Phys.\ Rev.\ D {\bf 86}, 103009 (2012); Phys.\ Rev.\ D {\bf 87}, no. 8, 083004 (2013);  
Class.\ Quant.\ Grav.\  {\bf 21}, 4209 (2004);  A.~R.~Liddle and S.~M.~Leach,  Phys.\ Rev.\ D {\bf 68}, 103503 (2003).

\bibitem{ref2}   P.~A.~R.~Ade {\it et al.} [Planck Collaboration], arXiv:1502.01594 [astro-ph.CO];
P.~A.~R.~Ade {\it et al.} [POLARBEAR Collaboration],  Phys.\ Rev.\ D {\bf 92}, 123509 (2015).

\bibitem{magnetized1}  M.~Giovannini,  Class.\ Quant.\ Grav.\  {\bf 23}, R1 (2006);  Phys.\ Rev.\ D {\bf 74}, 063002 (2006);  Phys.\ Rev.\ D {\bf 79}, 121302 (2009); Class.\ Quant.\ Grav.\  {\bf 30}, 205017 (2013).


\end{thebibliography}
\end{document}